\documentclass[preprint,jcp,floats,amsmath,amssymb]{revtex4}

\usepackage{graphicx,color,epstopdf}
\usepackage{mathtools}
\usepackage{subfigure}

\begin{document}
\title{Molecular finite-size effects in stochastic models of equilibrium chemical systems}

\author{Claudia Cianci}
\affiliation{School of Biological sciences, University of Edinburgh, Mayfield Road Edinburgh EH93JR Scotland, UK}
\author{Stephen Smith}
\affiliation{School of Biological sciences, University of Edinburgh, Mayfield Road Edinburgh EH93JR Scotland, UK}

\author{Ramon Grima}
\affiliation{School of Biological sciences, University of Edinburgh, Mayfield Road Edinburgh EH93JR Scotland, UK}

\begin{abstract}
The reaction-diffusion master equation (RDME) is a standard modelling approach for understanding stochastic and spatial chemical kinetics. An inherent assumption is that molecules  are point-like. Here we introduce the excluded volume reaction-diffusion master equation (vRDME) which takes into account volume exclusion effects on stochastic kinetics due to a finite molecular radius. We obtain an exact closed form solution of the RDME and of the vRDME for a general chemical system in equilibrium conditions. The difference between the two solutions increases with the ratio of molecular diameter to the compartment length scale.  We show that an increase in the fraction of excluded space can (i) lead to deviations from the classical inverse square root law for the noise-strength; (ii) flip the skewness of the probability distribution from right to left-skewed; (iii) shift the equilibrium of bimolecular reactions so that more product molecules are formed; (iv) strongly modulate the Fano factors and coefficients of variation. These volume exclusion effects are found to be particularly pronounced for chemical species not involved in chemical conservation laws. Finally we show that statistics obtained using the vRDME are in good agreement with those obtained from Brownian dynamics with excluded volume interactions. 
\end{abstract}
 

\maketitle

\vspace{0.8cm}

\section{Introduction}

A large number of studies have investigated the properties of noisy chemical dynamics (for recent reviews see for example \cite{Gillespie2007,Zhang2012}). The importance of the topic stems from an increasing interest in understanding the dynamics of chemical systems with small numbers of molecules for one or more species, for which stochastic effects are important. A natural example of such chemical systems are biochemical pathways inside cells \cite{GrimaSchnell2006a}; artificial examples include reactions occurring inside nano-spaces such as nano-reactors \cite{Vriezema2005} and carbon nanotubes \cite{Ugarte1996}.

The approaches at the heart of these studies include Brownian dynamics \cite{tenWolde2005,GillespieSeita2013}, the reaction-diffusion master equation (RDME) \cite{Baras1996,Isaacson2009} and its non-spatial counterpart, the chemical master equation (CME) \cite{Gillespie2007}. Brownian dynamics typically models point or hard spherical molecules which diffuse and interact with each other via chemical and steric interactions. The RDME provides an approximate spatially discretised version of Brownian dynamics, whereby space is divided into small volume elements (voxels), reactions occur between point molecules inside each voxel and diffusion of molecules is simulated by ``hopping'' of molecules between neighbouring voxels. The CME is a non-spatial approximation of the RDME, valid in the limit of well-mixed dynamics throughout the whole compartment. While Brownian dynamics is clearly the most realistic, the RDME and CME are far superior in terms of computational efficiency and have enabled the simulation of complex biochemical systems via the stochastic simulation algorithm (SSA) and its variants \cite{Gillespie2007,Bernstein2005}. Another advantage of master equations is that in many cases, they either can be solved exactly (see for example \cite{McQuarrie1964,Jahnke2007,Grima2012,GrimaSchnell2014}) or else their solution computed by means of an approximative method such as moment-closure approximation \cite{Ullah2009,GomezUribe2007,Schnoerr2014} or the system-size expansion \cite{ElfEhrenberg2003,Grima2014,ThomasPNAS2014,Thomas2015}) leading to insight which cannot be easily obtained by tediously long simulations using Brownian dynamics. 

Nevertheless a convincing argument can be made that the assumption of point molecules by the RDME and CME, is highly unrealistic, given that several experimental studies \cite{ZimmermannTrach1991,ZimmermannMinton1993,Zhou2008} have suggested that volume exclusion effects due to molecular crowding strongly modulate intracellular chemical equilibria and even play an important role in the regulation of gene expression rates \cite{Tan2013}. Brownian dynamics does not necessarily ignore such volume exclusion effects but is not an ideal simulation tool due to its heavy computational demand, not to mention its analytical impenetrability. A considerable number of studies have ignored chemical reaction kinetics and focused on understanding the diffusion of a tracer molecule in a sea of inert hard sphere molecules \cite{Saxton1994,Hofling2006,GrimaYalirakiBarahona2010,FanelliMcKane2010}. A few studies have, in contrast, sought to understand the effect of crowding on the stochastic chemical properties of very simple chemical systems in the reaction-limited regime, by renormalising the propensities of the CME to account for volume exclusion effects \cite{GillespiePetzold2007,Grima2010}. However to-date no general conclusions have been made, to our knowledge, about the impact of volume exclusion on the statistics of intrinsic noise in chemical systems. In other words, we would like to obtain insight into how the predictions of the RDME and CME for the distributions of molecule numbers of a general chemical system, are modified, if interacting molecules are modelled as hard particles with a finite radius.  

In this paper we take a step in this direction. We assume that all the molecules in a general chemical system are roughly of equal molecular size and devise a version of the RDME (the vRDME) which models reactions between such particles. Of course the assumption of a population of molecules with equal sizes is rough, however as we shall see it enables us to carry analytical calculations and to get a general idea of the impact of volume exclusion on the statistics of intrinsic noise. The paper is organised as follows. In Section II, we discuss in detail the RDME and the vRDME, and their non-spatial counterparts, the CME and vCME, pointing out their crucial differences. In Section III we use these master equations to derive exact closed-form expressions for the local and global distributions of molecule numbers in the presence and absence of volume exclusion effects. The relationship between the rate constants of the volume excluded and dilute approaches is discussed in Section IV. Next we use the results of Sections III and IV to explore the stochastic properties of chemical systems with no chemical conservation laws (Section V), with chemical conservation laws of a special type (Section VI) and with chemical conservation laws of a more general type (Section VII). The validity of the vRDME as an accurate approximation to a spatially continuous microscopic description is explored in Section VIII. We finally conclude by a summary and discussion in Section IX. 

\section{The CME, RDME, \lowercase{v}RDME,  and \lowercase{v}CME} \label{sec1}

In this section, we concisely describe the four mathematical frameworks used in this article: the CME, the RDME, and modified versions of these two, which take into account volume exclusion effects, and which we call the the vCME and vRDME respectively. To clarify the differences between the four mathematical frameworks we use the example of a simple reversible dimerisation whereby two molecules of a monomer (species A) diffuse and eventually bind to form a single molecule of the dimer (species B) and which at a later time dissociates back into the constituent monomers. 

\textbf{The CME} describes the stochastic time evolution of the molecule numbers of each chemical species in a well-mixed compartment. A major simplifying assumption is that the molecules are point particles. For the dimerisation reaction, the CME models the chemical process $A+A\xrightleftharpoons[k_1]{k_0} B$, where $k_0$ and $k_1$ are the rate constants for the forward and backward reactions. 

\textbf{The RDME}, is the spatial counterpart of the CME. The compartment is divided into $N$ subvolumes called voxels, each well-mixed (well-mixing is not assumed throughout the whole volume, only locally). The RDME describes the stochastic time evolution of the molecule numbers of each chemical species in each voxel, with the assumption that the particles are point-like. For the dimerisation reaction, the RDME models the processes:
\begin{align}\label{system11}
 A_i+A_i \xrightleftharpoons[k_1]{k_0} B_i,\quad A_i \xrightleftharpoons[k_D]{k_D} A_j,\quad B_i  \xrightleftharpoons[k_D]{k_D} B_j,\quad j \in Ne(i),
  \end{align}
where $A_i$ denotes species A in voxel $i$,  $B_i$ denotes species B in voxel $i$ and the notation $Ne(i)$ stands for the set of voxels which neighbour voxel $i$. The parameter $k_D$ has units of inverse time and is proportional to the diffusion coefficient $D$ of the species; specifically $k_D = D / \Delta x^2$ where $\Delta x$ is the side length of a voxel. The first reaction corresponds to the dimerisation reaction taking place in voxel $i$, while the second and third reactions model the diffusion of the monomer and the dimer between neighbouring boxes $i$ and $j$ with rate $k_D$. The RDME model with $4$ voxels is schematically represented in Fig.1(a). The particles are empty to underline that they occupy no volume, and the grid corresponds to the voxels. The relationship between the RDME and CME will be clarified further in the next section.

\textbf{The vRDME} is a modified version of the RDME, which we introduce in this paper as a means to take into account volume exclusion effects. In the vRDME, molecules are assumed to have roughly the same diameter and the voxel size is fixed to this length scale (unlike the RDME where the voxel size is arbitrary). A volume exclusion rule is implemented such that each voxel can accommodate at most one chemical molecule. An ``empty space species'' is defined whose molecule numbers reflect whether a voxel is empty or occupied. The volume exclusion rule is then implemented via the interaction of the empty space species and a chemical species. Bimolecular reactions involve the interaction of two chemical molecules in neighbouring voxels. For the dimerisation reaction, the vRDME models the processes:
\begin{align}\label{system2a}
 A_i+A_j \xrightleftharpoons[\tilde{k}_1]{\tilde{k}_0} B_i+E_j,\quad
  A_i+E_j \xrightleftharpoons[\tilde{k}_D]{\tilde{k}_D} E_i + A_j,\quad
  B_i+E_j \xrightleftharpoons[\tilde{k}_D]{\tilde{k}_D} E_i + B_j,\quad j \in Ne(i),  
\end{align}
where $E_i$ denotes an ``empty space molecule'' in voxel $i$ (the molecule number of species $E_i$ takes a value of zero if voxel $i$ is occupied and one if it is empty). The first process models the chemical reaction between two $A$ particles in neighbouring voxels and the other two processes model the diffusion of molecules between neighbouring voxels. Note that because we can interchange the indices $i$ and $j$, the chemical reaction between two $A$ particles in neighbouring voxels $i$ and $j$ leads to either a $B$ molecule in voxel $i$ or a $B$ molecule in voxel $j$. The reaction rates have a tilde to denote that these quantities are different than the rates used for the RDME (see later for the relationship between the rate constants of the RDME and the vRDME). The vRDME model with $N=36$ voxels is illustrated schematically in Fig.1(b). 

We note that the microscopic stochastic processes modelled by the vRDME have been previously simulated by means of Monte Carlo simulations on a two dimensional lattice, specifically applied to understanding diffusion-limited kinetics in crowded environments \cite{Berry2002,SchnellTurner2004,GrimaSchnell2006}. As well, the vRDME is a special case of a class of stochastic population models based on ``patch dynamics", a framework developed by McKane and Newman in the context of ecological systems \cite{McKaneNewman2004}. Specifically the vRDME corresponds to one of two types of spatial patch models, the case called ``micro-patch'' where each patch (each voxel in our terminology) can hold at most one individual. The bulk of studies to-date have however focused on the ``mesoscopic-patch" approach whereby each patch can hold at most a number $N$ of individuals where $N$ is typically a number much greater than one, and in which one assumes well-mixing and reactions occurring inside each patch, rather than between neighbouring patches (see for example \cite{Cianci2013}). 

\begin{figure} [h]
\label{fig1}
\centering
\subfigure[\ RDME with 4 voxels]{
\includegraphics[width=65mm]{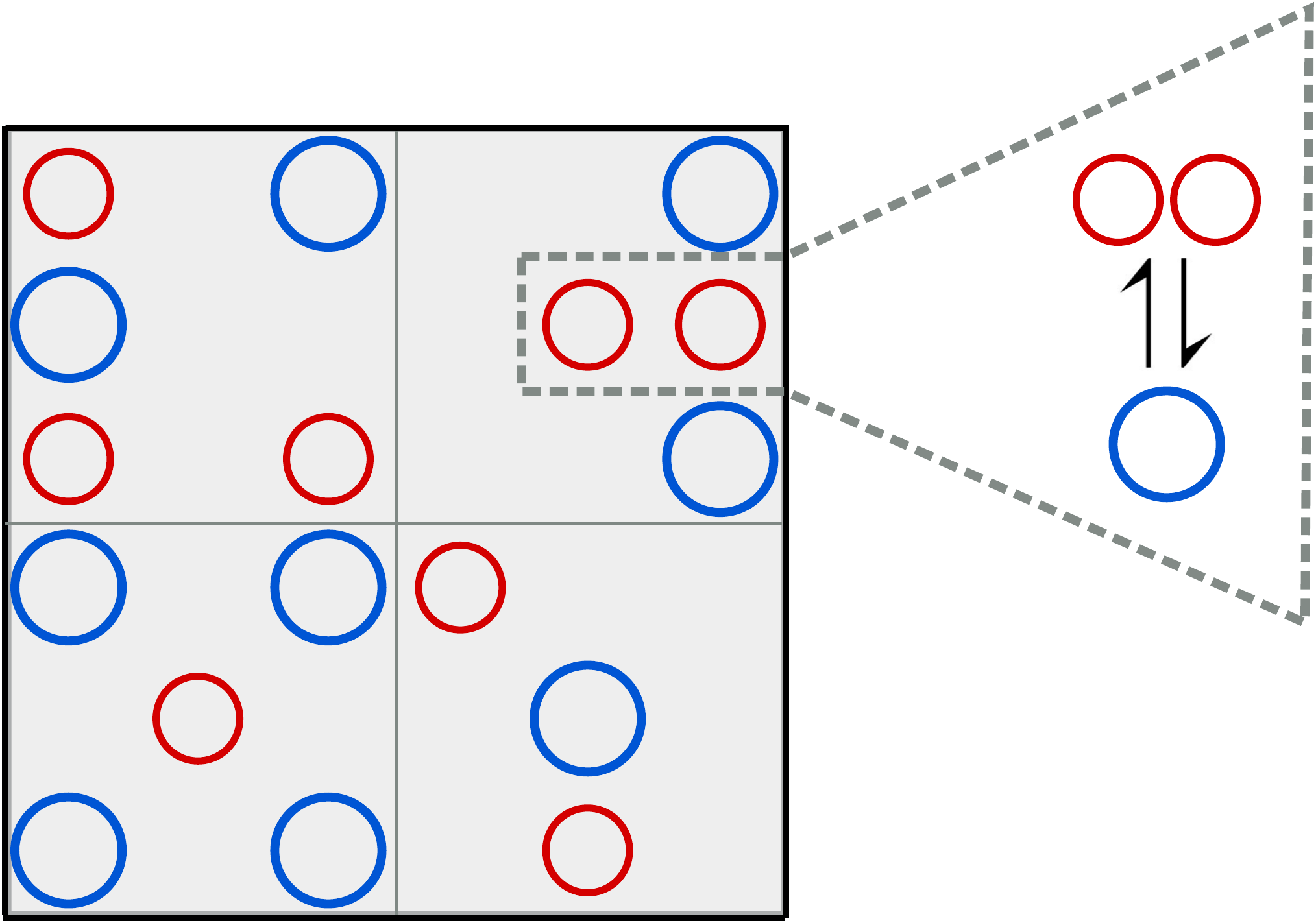}
}
\subfigure[\ vRDME with 36 voxels]{
\includegraphics[width=65mm]{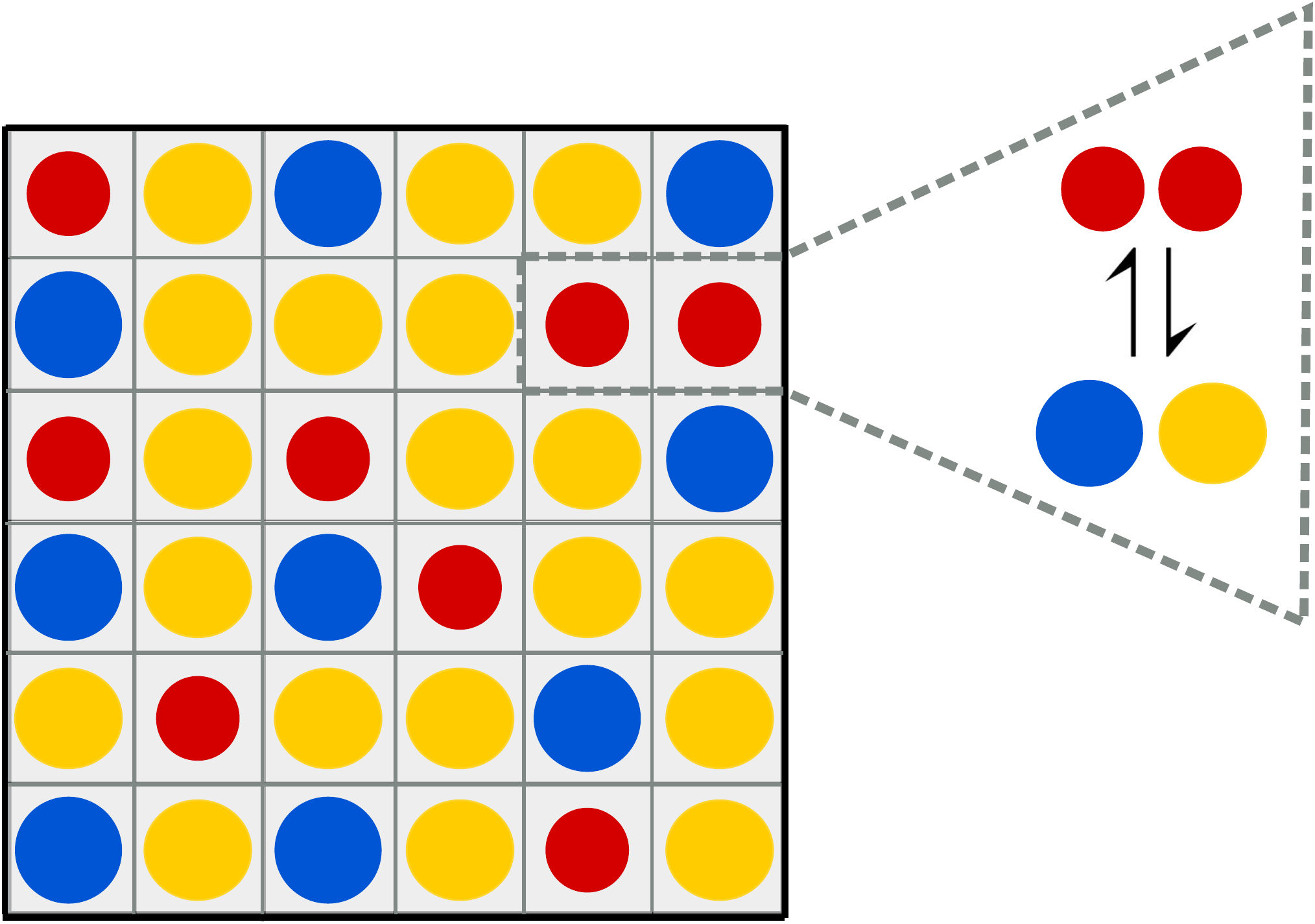}
}
\caption{Schematic illustrating the differences between the RDME and vRDME spatial modeling of the reaction $A+A\xrightleftharpoons[]{} B$. For the RDME (a), particles can react inside each of the 4 voxels and diffuse between neighbouring voxels. The red and blue circles denote particles of species $A$ and $B$ respectively. The particles are empty to denote that they occupy no volume (point particles) and can pass through each other. For the vRDME (b), the colour coding is the same except that we have also yellow circles denoting the ``empty space" $E$. Each voxel is occupied by at most one particle. Bimolecular reactions occur between neighbouring voxels. Diffusion involves the switching of an empty space molecule and a chemical molecule between two neighbouring sites. On the right of both (a) and (b), is an illustration of what happens when the dimerisation reaction occurs.}
 \end{figure}

\textbf{The vCME} is the non-spatial counterpart of the vRDME. The vCME is to the vRDME, what the CME is to the RDME. Hence the vCME is basically the CME but with two additional properties: (i) besides tracking the total number of molecules of each chemical species in the compartment, it also tracks the total number of empty space molecules in the compartment (this new non-chemical species is denoted as E); (ii) a global exclusion volume rule is imposed, namely that the total number of molecules of all species (chemical species and the empty space species) adds to a time-independent constant $N$ (which corresponds to the number of voxels in the vRDME). For the dimerisation reaction, the vCME models the processes: $A+A\xrightleftharpoons[\tilde{k}_1]{\tilde{k}_0} B+E$. Note the difference between the CME and vCME; the rate constants are also not the same, hence the tildes. The relationship between the vRDME and vCME will be clarified further in the next section.

We have in this section introduced the various mathematical frameworks by means of a simple chemical reaction system but these are applicable to more general systems of chemical reactions.

 \section{Exact solution of the CME, RDME, \lowercase{v}RDME and  \lowercase{v}CME in equilibrium conditions}\label{extension}
 
We will here focus on reversible chemical systems in equilibrium conditions, i.e., those in detailed balance \cite{vanKampen}. The reason for this restriction is that as we shall see, it enables us to write an explicit exact solution of the CME, RDME, vRDME and vCME, which will be crucial in later sections to understand the differences between them, i.e, the impact of molecular crowding on the stochastic dynamics of biochemical systems at the local (voxel) and global (compartment) level.  We shall start by summarising a result by van Kampen for the CME, which we will subsequently extend to the other three frameworks. 

\textbf{Global distribution of molecule numbers assuming point particle interactions}. Consider a well-mixed reversible system of $M$ chemical species interacting via $R$ chemical reactions where the $j^\text{th}$ reaction has the form:
\begin{align}\label{system1c}
s_{1j}X_{1}+...+s_{Mj}X_{M}& \xrightleftharpoons[k^{\prime}_j]{k_j} r_{1j}X_{1}+...+r_{Mj}X_{M},
\end{align}
where $X_i$ denotes the $i$th chemical species. Here $k_j$ and $k_j'$ are the rate constants for the forward and reverse reactions respectively and $r_{ij} - s_{ij}$ is the change in the number of molecules of species $X_i$ when reaction $j$ occurs.
We consider the system to be confined in a compartment of volume $\Omega$. The set of deterministic equilibrium constants \cite{physicalchembook} characterising this mass-action system are:
\begin{equation}
\label{deteqs}
 \phi_1^{r_{1j}-s_{1j}}\phi_2^{r_{2j}-s_{2j}}...\phi_M^{r_{Mj}-s_{Mj}} = \frac{k_j}{k_j'}, \quad j = 1, ..., R,
\end{equation} 
where $\phi_i$ is the deterministic concentration of species $X_i$ (as given by the conventional non-spatial rate equations). Furthermore we assume that the system has a number, $S$, of chemical conservation laws of the form:
\begin{equation}
\label{detclaws}
f_j( \vec{n}) = K_j, \quad j=1,...,S,
\end{equation}
where $\vec{n} = \{n_1, n_2, ..., n_M \}$ describes the number of molecules of each chemical species in the compartment of volume $\Omega$ and the $K_j$'s are time-independent constants set by the initial conditions and stoichiometry of the reaction system. Now the time-evolution of the global (whole compartment) distribution of molecule numbers assuming point particles and well-mixed conditions is given by the CME. Assuming mass-action kinetics, van Kampen showed that the exact equilibrium solution of the CME for system (\ref{system1c}) is given by \cite{vanKampen1976}:
\begin{equation}
\label{vKresult}
P(n_1, n_2, ..., n_M) = C \prod_{i=1}^M \frac{(\Omega \phi_i)^{n_i}}{n_i !} \prod_{j=1}^S \delta(f_j(\vec{n}), K_j),
\end{equation}
where $C$ is a normalisation constant, $\delta(.,.)$ is a Kronecker delta and $P(\vec{n})$ is the probability that the system is in state $\vec{n}$ in equilibrium. Hence the equilibrium solution is a product of Poisson distributions constrained by the chemical conservation laws. 

\textbf{Local distribution of molecule numbers assuming point particle interactions}. The result is also easily extensible to the RDME. The latter is a master equation describing the time-evolution of the probability that the system is in state $\{n_1^1,...,n_{M}^1,...,n_1^N,...,n_{M}^N\}$, where $n_i^j$ is the number of molecules of species $X_i$ in voxel $j$ and $N$ is the total number of voxels. This is a local description since it describes what happens at each point in space inside the compartment. Now at the voxel level, no chemical conservation laws hold; such laws are only global. For example, the reaction $X_1+X_1 \xrightleftharpoons[]{} X_2$ has the conservation law $n_1 + 2 n_2$ = constant, which is defined on the total number of molecules of $X_1$ and $X_2$ in the compartment, but locally in voxel $j$, $n_1^j + 2 n_2^j$ is not a constant due to the diffusive crosstalk with neighbouring voxels. It also follows that since we are considering a system in equilibrium, the deterministic concentration of a species in each voxel is the same as the deterministic concentration of the species in the whole compartment (that is equal to the solution of the non-spatial deterministic rate equations). Hence, given that there are only global conservation laws, that the local deterministic concentration is the same as the global deterministic concentration and that the voxel volume is $\Omega/N$, by analogy to the CME result above (Eq. (\ref{vKresult})), it follows immediately that the equilibrium probability distribution solution of the RDME is given by: 
\begin{align}
\label{RDMElocal}
P(n_1^1,...,n_{M}^1,...,n_1^N,...,&n_{M}^N) = C'  \prod_{k=1}^{N} \prod_{i=1}^{M}  \frac{((\Omega / N) \phi_i)^{n_i^k}}{n_i^k!}  \prod_{j=1}^S \delta(f_j(\vec{n}), K_j),
\end{align}
where $n_i$ is the global concentration of species $X_i$, i.e., $n_i = \sum_{j=1}^N n_i^j$.

\textbf{Global distribution of molecule numbers for finite size particle interactions}. The result of van Kampen can also be straightforwardly extended to the vCME. We will assume that the degree of molecular crowding is not so high that it prevents well-mixing in the limit of long times; this is the case if all molecules are mobile. The reactions here are modified than those for the CME because of the interaction of the chemical and empty space species. Hence the chemical system (\ref{system1c}) is now modified to:
\begin{equation}
\label{modsystem1c}
s_{1j}X_{1}+...+s_{M+1,j}X_{M+1}  \xrightleftharpoons[\tilde{k}^{\prime}_j]{\tilde{k}_j} r_{1j}X_{1}+...+r_{M+1,j}X_{M+1},
\end{equation}
where $X_i$, \ $i = 1,...,M$ are the chemical species and $X_{M+1}$ is the empty space species. The deterministic equilibrium constants are then given by:
\begin{equation}
\label{deteqsC}
 \tilde{\phi}_1^{r_{1j}-s_{1j}}\tilde{\phi}_2^{r_{2j}-s_{2j}}...\tilde{\phi}_{M+1}^{r_{M+1,j}-s_{M+1,j}} = \frac{\tilde{k}_j}{\tilde{k}_j'}, \quad j = 1, ..., R,
\end{equation} 
where $\tilde{\phi}_i$ is the deterministic concentration of species $X_i$ according to the deterministic rate equations one would write for the reaction scheme (\ref{modsystem1c}). Another difference from the CME is that besides the $S$ chemical conservation laws given by Eq. (\ref{detclaws}), we now also have an additional global conservation law stemming from volume exclusion, namely 
\begin{equation}
\label{detclawsC}
 \sum_{i=1}^{M+1} n_i = N,
\end{equation}
where $N$ is the total number of molecules which can be maximally fit in the compartment. Given this information, by analogy with the CME result above (Eq. (\ref{vKresult})), it follows immediately that the equilibrium probability distribution solution of the vCME is given by:
\begin{equation}
\label{vCMEsol}
P(n_1,...,n_{M+1}) = C'' \prod_{j=1}^{M+1} \frac{(\Omega \tilde{\phi}_j)^{n_j}}{n_j!} \delta(\sum_{i=1}^{M+1} n_i, N) \prod_{k=1}^S \delta(f_k(\vec{n}), K_k).
\end{equation}
Note that the global distribution is explicitly a function of $N$; this is in contrast to the global distribution of the CME which has no such information.

\textbf{Local distribution of molecule numbers for finite size particle interactions}. The result of van Kampen can also be extended to the vRDME. We will assume that molecular crowding does not prevent diffusion between any two voxels in the compartment; this is the case if all molecules are mobile. This requirement is needed to satisfy detailed balance.
Since the system is in equilibrium, the deterministic concentrations in each voxel are the same as the deterministic concentrations in the whole compartment according to the vCME. The state vector is $\{n_1^1,...,n_{M+1}^1,...,n_1^N,...,n_{M+1}^N\}$, where $n_{i}^j$ is the number of molecules of species $X_i$ in voxel $j$ ($1 \le i \le M$), $n_{M+1}^j$ is the number of empty space molecules in voxel $j$ and $N$ is the total number of voxels. A crucial difference from the RDME is that in addition to global conservation laws, now we also have a conservation law in each voxel, namely there can be at most one molecule in each voxel, i.e., $\sum_{k=1}^{M+1} n_k^i = 1$, for $i=1,..,N$. Given this information and taking into account the fact that the voxel volume is $\Omega/N$, by analogy with the CME result above (Eq. (\ref{vKresult})), it follows immediately that the equilibrium probability distribution solution of the vRDME is given by: 
\begin{align}
\label{vRDMElocal}
P(n_1^1,...,n_{M+1}^1,...,n_1^N,...,&n_{M+1}^N) = C'''  \prod_{k=1}^{N} \prod_{i=1}^{M+1}  \frac{((\Omega / N) \tilde{\phi}_i)^{n_i^k}}{n_i^k!} \delta(\sum_{i=1}^{M+1} n_i^k, 1) \prod_{j=1}^S \delta(f_j(\vec{n}), K_j).
\end{align}

Note that because of the constraints due to conservation laws (chemical or volume exclusion), generally the mean concentration vector calculated using the exact equilibrium solutions of the CME, RDME, vCME and vRDME are not equal to their deterministic concentration vector ($\vec{\phi}$ and $\vec{\tilde{\phi}}$) respectively, except in the macroscopic limit of large volumes.

Note also that the local distribution solutions are independent of the underlying connectivity of the lattice of the RDME and vRDME. This is because in equilibrium, the solution of a master equation is generally a product of Poissonians constrained by local and global conservation laws \cite{vanKampen1976}, and these laws are not in any way influenced by the lattice connectivity. Of course as previously mentioned the condition of detailed balance (equilibrium) is compatible only with those lattices such that there exists a path connecting any two lattice points. Thus this is the only requirement on a lattice, for our results to hold.

It is also a fact that in detailed balance (equilibrium) conditions, the global probability distribution calculated starting from the local distribution solution of the RDME exactly matches the global distribution solution of the CME, independent of the diffusion coefficients. The same applies to the vRDME and the vCME. This maybe intuitive to some readers but for the sake of completeness we provide a proof in Appendix A. Thus although we initially presented the vCME in Section II via an intuitive approach, the macroscopic solution of the vCME stands on a solid basis since it can be obtained from the microscopic approach of the vRDME. 

The rest of this article is devoted to obtaining insight into the effect of volume exclusion on the global distribution of molecule numbers and to a much lesser extent on the local distribution of molecule numbers. Due to the equivalence of the vRDME and vCME in equilibrium conditions and at the global level of description, we shall use both interchangeably when referring to any conclusions made assuming a finite molecular radius. Similarly we shall use RDME and CME interchangeably when discussing conclusions made assuming point particles.
  
\section{Relationship of rate constants in the CME and \lowercase{v}CME} \label{detsec}

Previously we have denoted the rate constants in the vCME formalism by tildes to clarify that they are different to those in the CME. Here we show the connection between the two. 

We start by noting that the dilute limit of infinitesimally small molecules corresponds to the limit of infinitely large number of voxels $N$ (in the vRDME and vCME) at constant compartment volume $\Omega$. Equivalently this corresponds to the limit, i.e., $\Omega \tilde{\phi}_{M+1} \rightarrow N$, where practically all of space is empty (species $X_{M+1}$ is the empty space species). In this limit, the deterministic (global) concentrations of the the vCME and of the CME must be equal, which for system (\ref{system1c}) implies:
\begin{align}
\lim_{\tilde{\phi}_{M+1} \rightarrow N/\Omega} \tilde{\phi}_1^{r_{1j}-s_{1j}} \tilde{\phi}_2^{r_{2j}-s_{2j}}...\tilde{\phi}_M^{r_{Mj}-s_{Mj}} = \phi_1^{r_{1j}-s_{1j}} \phi_2^{r_{2j}-s_{2j}}...\phi_M^{r_{Mj}-s_{Mj}}, \quad j = 1, ..., R.
\end{align}
This statement together with Eqs. (\ref{deteqs}) and (\ref{deteqsC}) implies:
\begin{align}
\label{relCNC}
\frac{\tilde{k}_j'}{\tilde{k}_j} \biggl(\frac{N}{\Omega}\biggr)^{r_{M+1,j}-s_{M+1,j}} = \frac{k_j'}{k_j},  \quad j = 1, ..., R.
\end{align}
This equation encapsulates the relationship between the rate constants of the volume excluded and dilute probabilistic descriptions. For example for the reversible dimerisation reaction previously considered, the CME and vCME formulations model the reactions $X_1+X_1 \xrightleftharpoons[k_1']{k_1} X_2$ and $X_1+X_1 \xrightleftharpoons[\tilde{k}_1']{\tilde{k}_1} X_2 + X_3$ respectively (where $X_3$ is the empty space species), which implies the relation 
$\tilde{k}_1'/\tilde{k}_1 = k_1' \Omega/k_1 N$.  

It can be shown using the model reduction technique developed in \cite{Smith2015b} that in the limit of abundant empty space species (the dilute limit), the global distribution over the molecule numbers of the chemical species as given by the vCME, Eq. (\ref{vCMEsol}), tends to the global distribution of the CME, Eq. (\ref{vKresult}).

Using the relationship between the rate constants derived above, we can also understand how the effective equilibrium constant changes as a function of the strength of volume exclusion effects. According to the standard definition in physical chemistry and thermodynamics \cite{physicalchembook}, the effective equilibrium constant of the jth reaction in system (\ref{system1c}) in volume excluded and dilute conditions are respectively given by:
\begin{align}
\tilde{E}_j&=\tilde{\phi}_1^{r_{1j}-s_{1j}} \tilde{\phi}_2^{r_{2j}-s_{2j}}...\tilde{\phi}_M^{r_{Mj}-s_{Mj}}, \\
E_j&=\phi_1^{r_{1j}-s_{1j}} \phi_2^{r_{2j}-s_{2j}}...\phi_M^{r_{Mj}-s_{Mj}}.
\end{align}
Now by Eq. (\ref{deteqsC}) and Eq. (\ref{relCNC}) we then have:
\begin{align}
\tilde{E}_j&= \frac{\tilde{k}_j}{\tilde{k}_j' \tilde{\phi}_{M+1}^{r_{M+1,j}-s_{M+1,j}}} = \frac{k_j}{k_j'} \biggl(\frac{N}{\Omega \tilde{\phi}_{M+1}} \biggr)^{r_{M+1,j}-s_{M+1,j}} \\&=\frac{E_j}{(\mathrm{fraction \ of \ empty \ space})^{r_{M+1,j}-s_{M+1,j}}}.
\end{align}

This implies that the effective equilibrium constant of unimolecular reactions is unaffected by crowding since in this case $r_{M+1,j}=s_{M+1,j}=0$ because no space is involved. The effective equilibrium constant for bimolecular reactions is however increased relative to the one for non-crowded conditions, $\tilde{E}_j > E_j$, since in this case a single molecule of empty space is produced when two molecules bind to form a single molecule ($r_{M+1,j} = 1,s_{M+1,j} = 0$).  

This result for bimolecular reactions can indeed be deduced without any calculation but with the application of Le Chatelier's principle in physical chemistry \cite{physicalchembook} to the vCME formalism. This principle states that a system in equilibrium will counteract the effect of an applied change by adjusting to a new equilibrium. Now if we consider the reaction $X_1+X_1 \xrightleftharpoons[]{} X_2$, this is modelled in the vCME formalism as $X_1+X_1 \xrightleftharpoons[]{} X_2 + X_3$ and hence by Le Chatelier's principle, an increase in volume exclusion, i.e., a decrease in $X_3$ (the empty space species) will induce the system to shift its equilibrium to the right to counteract this decrease, in the process causing an increase in the amount of species $X_2$ and a decrease in the amount of species $X_1$ which amounts to an increase in the effective equilibrium constant. 

These results for unimolecular and bimolecular reactions encapsulate the essence of the thermodynamic theory of crowding developed by Minton and co-workers \cite{Zhou2008}. However it is the first time, to our knowledge, that they have been obtained using the deterministic limit of a master equation description. 

\section{Stochastic description of chemical systems without chemical conservation laws}

In this section we use the results of Sections \ref{extension} and \ref{detsec} to show that {\emph{if there are no chemical conservation laws then the marginal distribution of the global molecule numbers of each chemical species $X_i$ is Poisson ($\Omega \phi_i$) if crowding is ignored and Binomial ($N, \Omega \tilde{\phi}_i/N$) if crowding is taken into account}}. Here $\Omega$ is the compartment volume, $N$ is the maximum number of particles which can be placed in the compartment if volume exclusion is taken into account, and $\phi_i$ and $\tilde{\phi}_i$ are the deterministic concentrations of the CME and vCME respectively. We shall call this Statement 1. In what follows, we discuss the physical implications of this statement, as well as confirm our results using stochastic simulations of the CME and the vCME applied to an open homodimerisation reaction. 

\subsection{Derivation of Statement 1}

As shown in section \ref{extension}, the global probability distribution assuming point particles, for a system with $M$ chemical species, is generally given by the solution of the CME, namely Eq. (\ref{vKresult}). Now if there are no chemical conservation laws, i.e., there is no factor $\delta(f_j(n_1,n_2,...,n_M), K_j)$ in Eq. (\ref{vKresult}), then the  global solution is simply a multivariate Poisson distribution:
\begin{equation}\label{solution1}
P(\vec{n})=e^{-\sum_{i=1}^M (\Omega\phi_i)} \frac{(\Omega\phi_{1})^{n_1}}{n_1!}
\frac{(\Omega\phi_{2})^{n_2}}{n_2!}\frac{(\Omega\phi_{3})^{n_3}}{n_3!}...\frac{(\Omega\phi_{M})^{n_M}}{n_M!},
\end{equation}
and hence it follows that the marginal distribution of each chemical species $X_i$ when volume exclusion is ignored, is a Poisson with mean $\Omega \phi_i$. 

We also showed that the global probability distribution for molecules with a finite radius and assuming $N$ of them can at most be packed in the compartment, for a system with $M$ chemical species (and an additional species $X_{M+1}$ representing free space), is generally given by the solution of the vCME, namely Eq. (\ref{vCMEsol}). Now if there are no chemical conservation laws, i.e., there are no factors of the type $\delta(f_k(n_1,n_2,...,n_M), K_k)$ in Eq. (\ref{vCMEsol}), then the normalised global probability distribution is given by:
\begin{equation}\label{multinomial}
P(\vec{n})=N!\frac{(\Omega \tilde{\phi}_1 / N)^{n_1}}{n_1!}\frac{(\Omega \tilde{\phi}_2/N)^{n_2}}{n_2!}\frac{(\Omega\tilde{\phi}_3/N)^{n_3}}{n_3!}...\frac{(\Omega\tilde{\phi}_{M+1}/N)^{n_{M+1}}}{n_{M+1}!}\delta \left( n_1+...+n_{M+1},N\right),
\end{equation}
which is a Multinomial distribution with parameters $(\{\Omega \tilde{\phi}_1 / N,...,\Omega \tilde{\phi}_{M+1}/ N \},N)$. Note that $\Omega \tilde{\phi}_i/N$ is the fraction of space occupied by particles of species $X_i$ and consequently $\sum_{i=1}^{M+1} \Omega \tilde{\phi}_i/N = 1$. It is well known that the marginal distributions of a multinomial distribution are Binomial \cite{Binomref}. For species $X_i$, the marginal distribution is thus Binomial with parameters $(N, \Omega \tilde{\phi}_i/N)$:
\begin{equation}
 P(n_i)=N! \frac{(\Omega{\tilde\phi}_{i}/N)^{n_i}}{n_i! (N-n_i)!}
\biggl(1 - \frac{\Omega \tilde{\phi}_{i}}{N}\biggr)^{N-n_i},\quad i=1,...,M.
\end{equation}

\subsection{Dilute limit} 

Consider the dilute limit $\Omega \tilde{\phi}_{M+1} \rightarrow N$. This can equivalently be seen as the limit of large numbers of voxels (at constant compartment volume $\Omega$) in the vRDME such that the occupied volume fractions of all chemical species (except the empty space species) tend to zero and the deterministic solution of the vRDME (vCME) approaches that of the RDME (CME), i.e.,  $N \rightarrow \infty$ and $\Omega \tilde{\phi}_i / N \rightarrow 0$, such that $\Omega \tilde{\phi}_i \rightarrow {\Omega \phi}_i$ for $1 \le i \le M$. Note that the last limit $\Omega \tilde{\phi}_i \rightarrow {\Omega \phi}_i$ for $1 \le i \le M$ follows by the specific relationship between the rate constants of the vRDME and of the RDME enforced in Section (\ref{detsec}). Note also that the dilute limit implies infinitesimally small molecules, since the volume of each molecule is roughly that of a voxel $\Omega / N$.  It then follows by the Poisson limit theorem \cite{Papolis}, that in the dilute limit, the global marginal distribution of the vRDME, Binomial $(N, \Omega \tilde{\phi}_i/N)$, tends to the global marginal distribution of the RDME, Poisson $(\Omega \phi_i)$. 

\subsection{Statistical measures and Physical implications} 

The Fano factor ($F$) is defined as the ratio of the variance and the mean, and is a measure of how much a distribution differs from a Poisson distribution; the coefficient of variation ($CV$) is defined as the ratio of the standard deviation and the mean, and is a measure of how ``noisy" a system is; and the skewness ($SK$) of a distribution is the third standardised moment of the distribution, and is a measure of how asymmetrical it is. These measures are well known for the Poisson and Binomial distributions and hence we can state that assuming point particles, the statistics of chemical species $X_i$ are given by:
\begin{align}
\label{statsP}
\langle n_i\rangle&=\Omega\phi_i, \quad F_i=1, \quad CV_i^2=\frac{1}{\langle n_i \rangle}, \quad SK_i=\frac{1}{\sqrt{\langle n_i \rangle}},
\end{align}
while for those modelled assuming a finite molecular radius, the statistics of chemical species $X_i$ are given by:
\begin{align}
\label{cqts}
\langle n_i\rangle&=\Omega \tilde{\phi}_i, \quad &F_i&=1-\frac{\langle n_i \rangle}{N}, \nonumber \\ ~CV_i^2&=\frac{1-\frac{\langle n_i \rangle}{N}}{\langle n_i\rangle}, \quad &SK_i&=\frac{1-2\frac{\langle n_i \rangle}{N}}{\sqrt{\langle n_i\rangle (1-\frac{\langle n_i \rangle}{N})}}.
\end{align}
The differences between equations Eqs. (\ref{statsP}) and Eq. (\ref{cqts}) are illustrated in Fig. 2 where we plot the qualitative behaviour of the Fano factor, the coefficient of variation squared and the skewness for a system, in which volume exclusion effects are neglected due to the assumption of point particles (green lines) and when they are taken into account (red lines). 

\begin{figure} [h]
\centering
\subfigure[]{
\includegraphics[width=50mm]{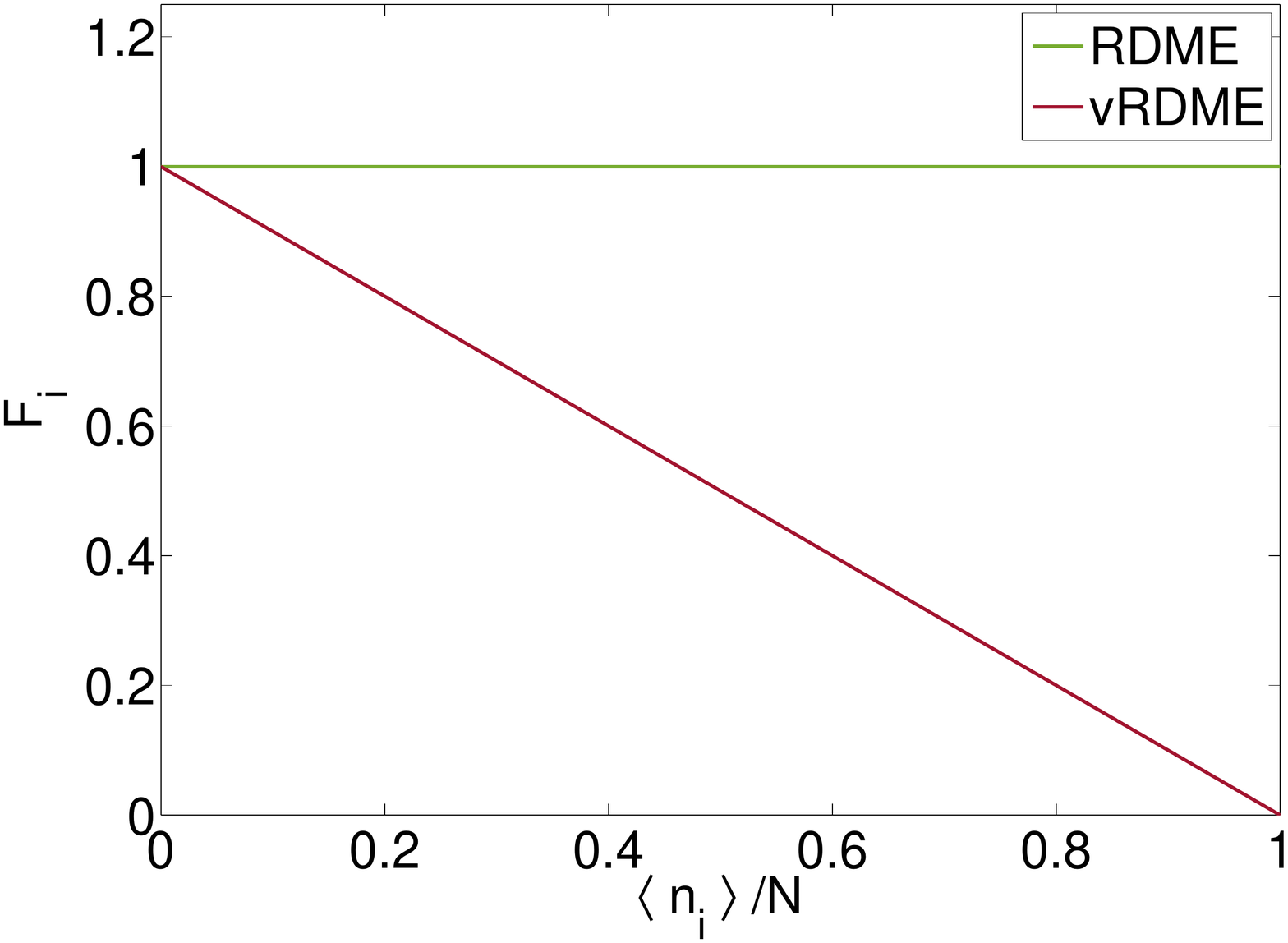}
\label{fig:subfig1}
}
\subfigure[]{
\includegraphics[width=50mm]{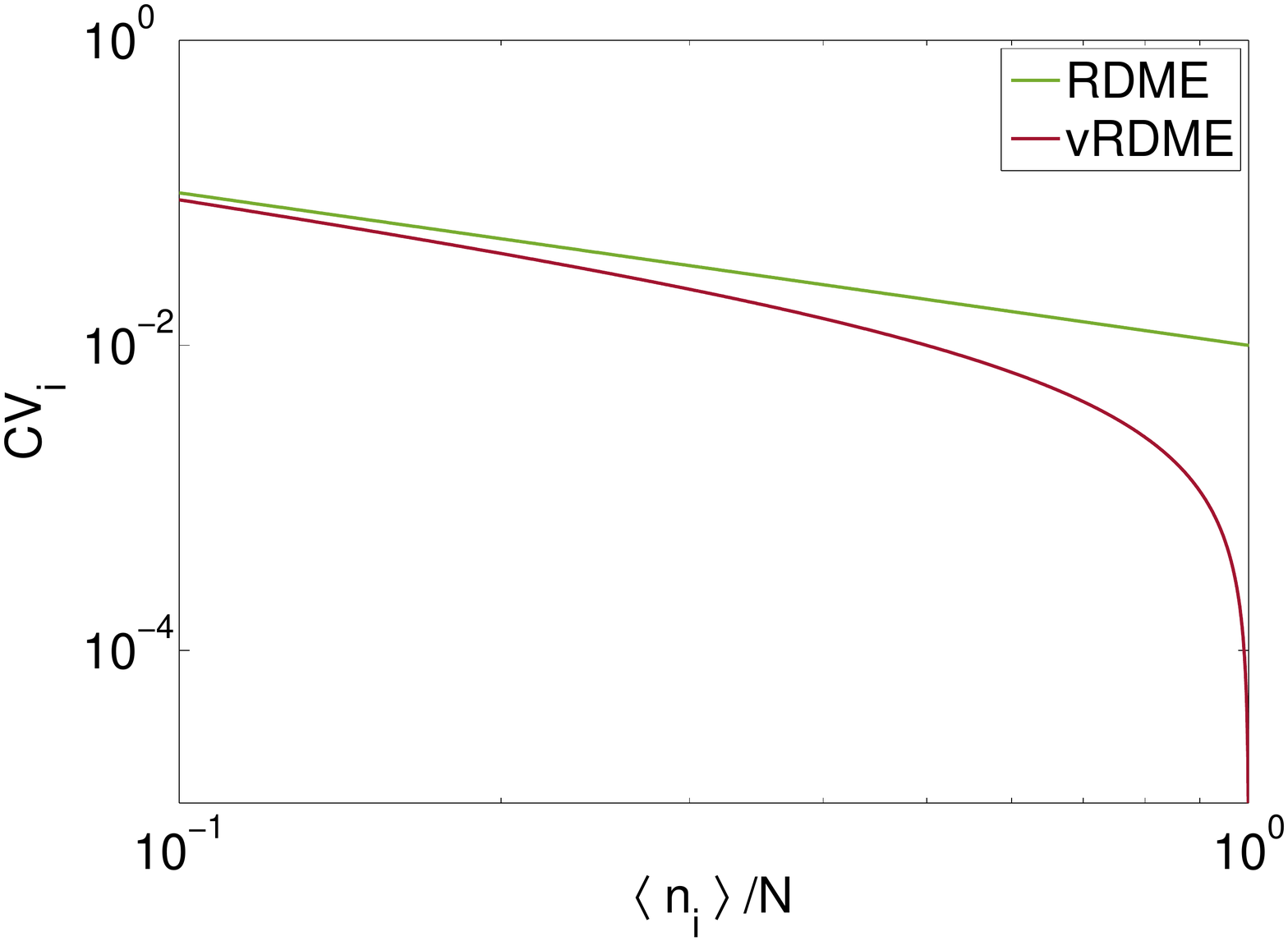}
\label{fig:subfig2}
}
\subfigure[]{
\includegraphics[width=50mm]{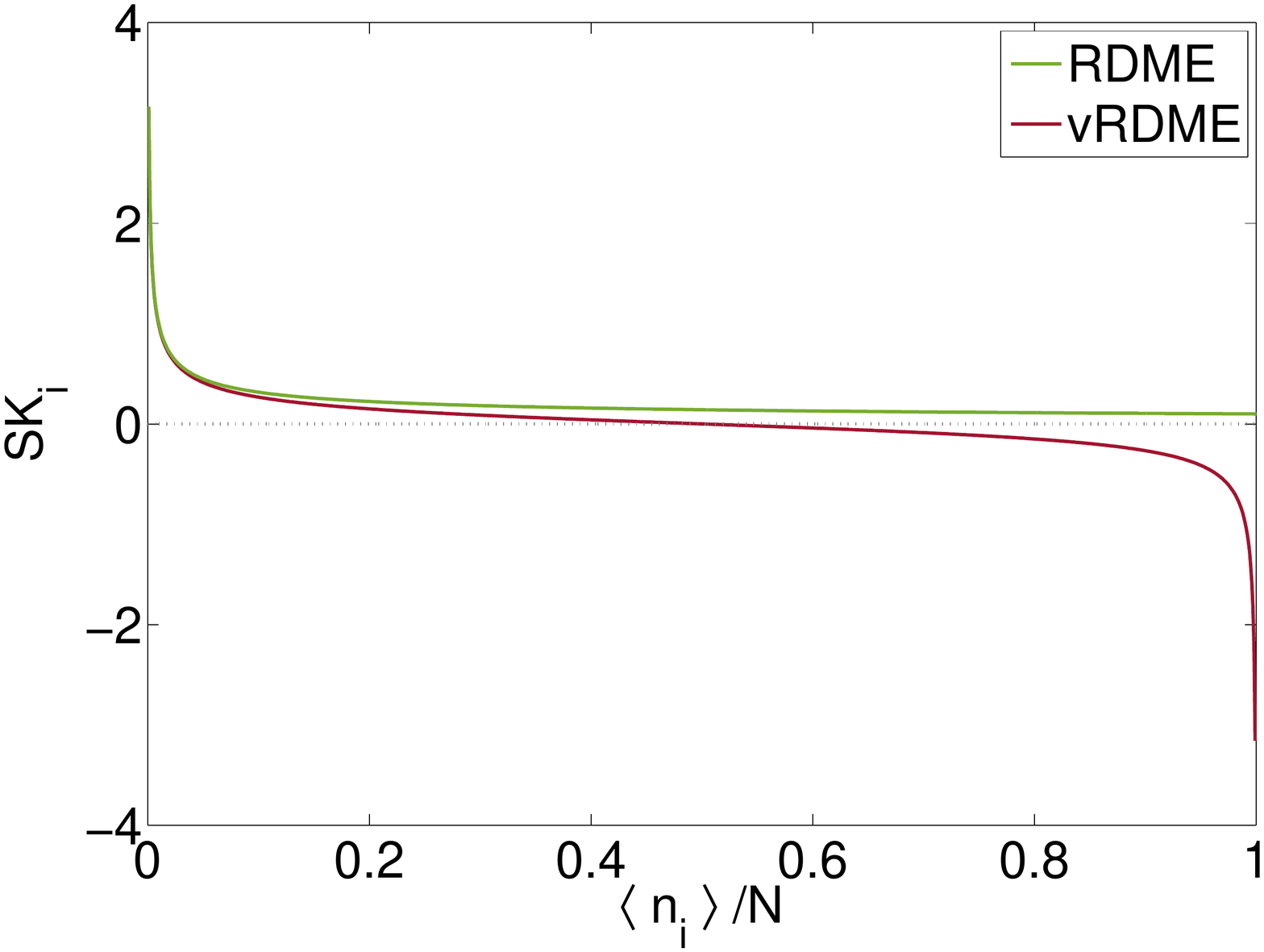}
\label{fig:subfig1}
}
\caption{The Fano factor (a), coefficient of variation (b), and skewness (c) of a species $X_i$ in a chemical system without chemical conservation laws. The red and green lines correspond to the statistics predicted by assuming a finite molecular radius (as given by Eqs. (\ref{cqts})) and by assuming point particles (as given by Eqs. (\ref{statsP})), respectively. Volume exclusion effects become more appreciable as the occupied volume fraction of space tends to unity ($\langle n_i \rangle/N \rightarrow 1$) which causes the Fano factor to decrease (a), deviations from the inverse square root law for noise strength (b) and the skewness to switch from positive to negative (c). }
 \end{figure}

The physical implication of these results is as follows. {\emph{As the fraction of occupied space increases, i.e, as $\langle n_i \rangle / N \rightarrow 1$, the fluctuations change from Poissonian ($F = 1$) to sub-Poissonian ($F < 1$), the well known classical noise-strength power law \cite{vanKampen,BarEven2006} ($CV_i \propto \langle n_i \rangle^{-1/2}$ ) becomes invalid and the distribution of fluctuations changes from being skewed to the right (positive skewness) to being skewed to the left (negative skewness). The latter occurs when the occupied volume fraction $\langle n_i \rangle / N$ exceeds $1/2$}}. 

Another interpretation of the results, is that if one ignores volume exclusion effects, i.e., employs the CME/RDME to model chemical systems without chemical conservation laws, then the dependence of the Fano factor, coefficient of variation and the skewness on the mean molecule numbers is qualitatively wrong for high molecule numbers. It also follows from the properties of multivariate Poisson and multinomial distributions that ignoring volume exclusion implies zero covariance between the molecule numbers of different species while taking it into account implies a negative covariance.

\subsection{Application: open homodimerisation reaction}

We consider the dilute (point particle) chemical system:
\begin{align}\label{system7n}
\emptyset  \xrightleftharpoons[k_1]{k_0} X_1,\quad
X_1+X_1 \xrightleftharpoons[k_3]{k_2}X_2,
\end{align}
whereby a species $X_1$ is produced and subsequently two molecules of this species reversibly bind to form another molecule of type $X_2$. This system follows no chemical conservation laws and hence is of the type discussed above. The Fano factor, coefficient of variation and skewness for the fluctuations in both species are given by Eqs. (\ref{statsP}) together with the deterministic equilibrium constants:
\begin{equation}
\label{eqcsts1}
\phi_1= \frac{k_0}{k_1}, \quad \frac{\phi_2}{\phi_1^2}=\frac{k_2}{k_3}.
\end{equation}
This procedure leads to the following equations:
\begin{equation}
\label{statsapp1a}
F_1=1,~F_2=1,~CV_1^2=\frac{k_1}{\Omega k_0},~CV_2^2=\frac{k_3k_1^2}{k_2k_0^2\Omega},~SK_1=\sqrt{\frac{k_1}{\Omega k_0}},~SK_2=\sqrt{\frac{k_3k_1^2}{k_2k_0^2\Omega}}.
\end{equation}

The volume exclusion version (assuming a finite molecular radius) of the chemical system (\ref{system7n}) is given by:
 \begin{align}\label{system}
X_3  \xrightleftharpoons[\tilde{k}_1]{\tilde{k}_0} X_1,\quad
 X_1+X_1 \xrightleftharpoons[\tilde{k}_3]{\tilde{k}_2}X_2+X_3,
 \end{align}
where species $X_3$ is the empty space species. The Fano factor, coefficient of variation and skewness for the fluctuations in both species are given by Eqs. (\ref{cqts}) together with the deterministic equilibrium constants:
\begin{equation}
\label{eqcsts2}
\frac{\tilde{\phi}_1}{\tilde{\phi}_3}= \frac{\tilde{k}_0}{\tilde{k}_1}, \quad \frac{\tilde{\phi}_2 \tilde{\phi}_3}{\tilde{\phi}_1^2}=\frac{\tilde{k}_2}{\tilde{k}_3},
\end{equation}
and the conservation law due to volume exclusion:
\begin{equation}
\tilde{\phi}_1 + \tilde{\phi}_2 + \tilde{\phi}_3 = N / \Omega,
\end{equation}
where $N$ is the total number of molecules which can be contained in the compartment. Furthermore we know that in the dilute limit, $\phi_3 \rightarrow N / \Omega$, the effective equilibrium constants of the crowded system must equal the equilibrium constants of the non-crowded system (as previously discussed at length in Section {\ref{detsec}}). Thus using Eqs. (\ref{eqcsts1}) and (\ref{eqcsts2}), we have the following relationship between the rate constants of the crowded system and of the non-crowded system:
\begin{equation}
\frac{\tilde{k}_0 N}{\tilde{k}_1 \Omega} = \frac{k_0}{k_1},  \quad \frac{\tilde{k}_2 \Omega}{\tilde{k}_3 N} = \frac{k_2}{k_3}.
\end{equation}

The overall procedure described above leads to the following equations:
\begin{align}
\label{statsapp1b}
F_1&=\frac{k_1^2 k_3 N + k_0^2 k_2 \Omega}{k_1^2 k_3 N + k_0 (k_0 k_2 + k_1 k_3) \Omega},~&F_2&=\frac{k_1 k_3 (k_1 N + k_0 \Omega)}{k_1^2 k_3 N + k_0 (k_0 k_2 + k_1 k_3) \Omega},\nonumber \\~CV_1^2&=\frac{k_1^2 k_3 N + k_0^2 k_2 \Omega}{k_0 k_1 k_3 N \Omega}, ~&CV_2^2&=\frac{k_1^2 k_3 N + k_0 k_1 k_3 \Omega}{k_0^2 k_2 N \Omega},\nonumber \\~SK_1&={\frac{k_0^2 k_2 \Omega + k_1 k_3 (k_1 N - k_0 \Omega)}{\sqrt{k_0 k_1 k_3 N \Omega (k_1^2 k_3 N + k_0^2 k_2 \Omega)}}},~&SK_2&=\frac{k_1^2 k_3 N + k_0 (k_1 k_3 -k_0 k_2) \Omega}{k_0 \sqrt{k_1 k_2 k_3 N \Omega (k_1 N + k_0 \Omega)}}.
\end{align}

Comparing the statistical quantities Eqs. (\ref{statsapp1a}) and (\ref{statsapp1b}), one notices the stark difference in the parametric dependence of these quantities if volume exclusion effects are taken into account. These differences are illustrated in Fig. 3 where the solid lines  show the analytical predictions for the Fano factor, coefficient of variation, and skewness of both species as a function of the parameter $k_0$, for dilute (upper panel) and volume exclusion conditions (lower panel). The analytical formulae are compared with data from the Stochastic Simulation Algorithm (SSA, open circles) sampling the CME and the vCME, as evidence of their exactness. 

As one can appreciate from these plots, the dependence on $k_0$ is strongly affected by volume exclusion, except of course in the limit of small $k_0$ where there are few particles in the compartment. Some qualitative differences which are particularly noticeable and interesting are: (i) volume exclusion has very little impact on the Fano factor of species $X_1$ but a strong impact on the Fano factor of species $X_2$ (a change from constant to strongly monotonic decreasing as a function of $k_0$); (ii) in contrast, volume exclusion has a strong  impact on the coefficient of variation of species $X_1$ (a change from a monotonic decreasing function to a parabolic function of $k_0$) but little impact on the coefficient of variation of $X_2$; (iii) the skewness of species $X_2$ becomes negative as $k_0$ increases beyond a certain threshold, for volume excluded conditions, but remains positive in dilute conditions. These stark differences suggest that the parameter dependencies of various statistical quantities that one obtains using conventional dilute approaches (the RDME and CME), may not always reflect the actual parameter dependencies \emph{in vivo}.   

\begin{figure} [h]
\label{fig3n}
\centering
\subfigure[]{
\includegraphics[width=50mm]{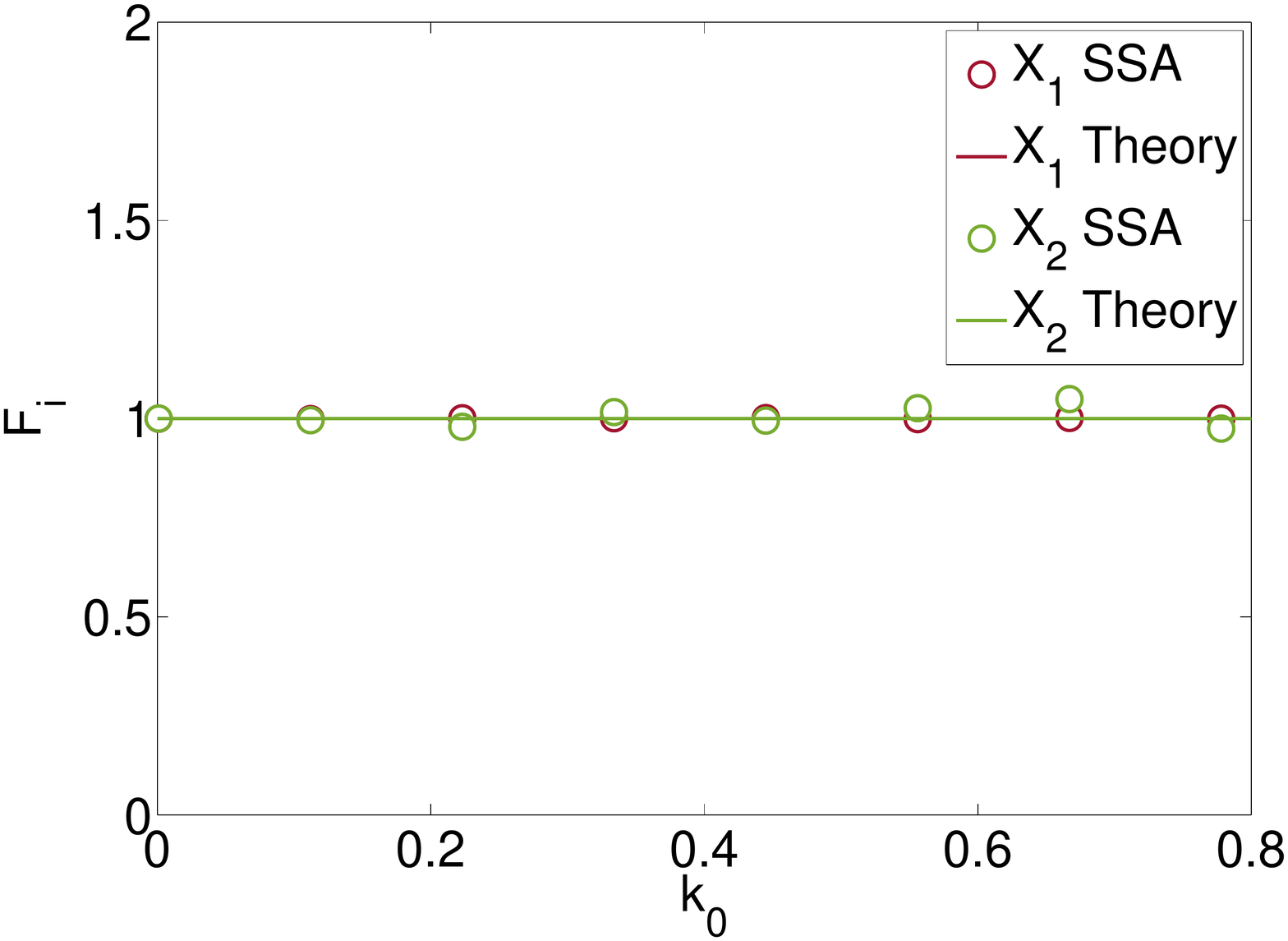}
}
\subfigure[]{
\includegraphics[width=50mm]{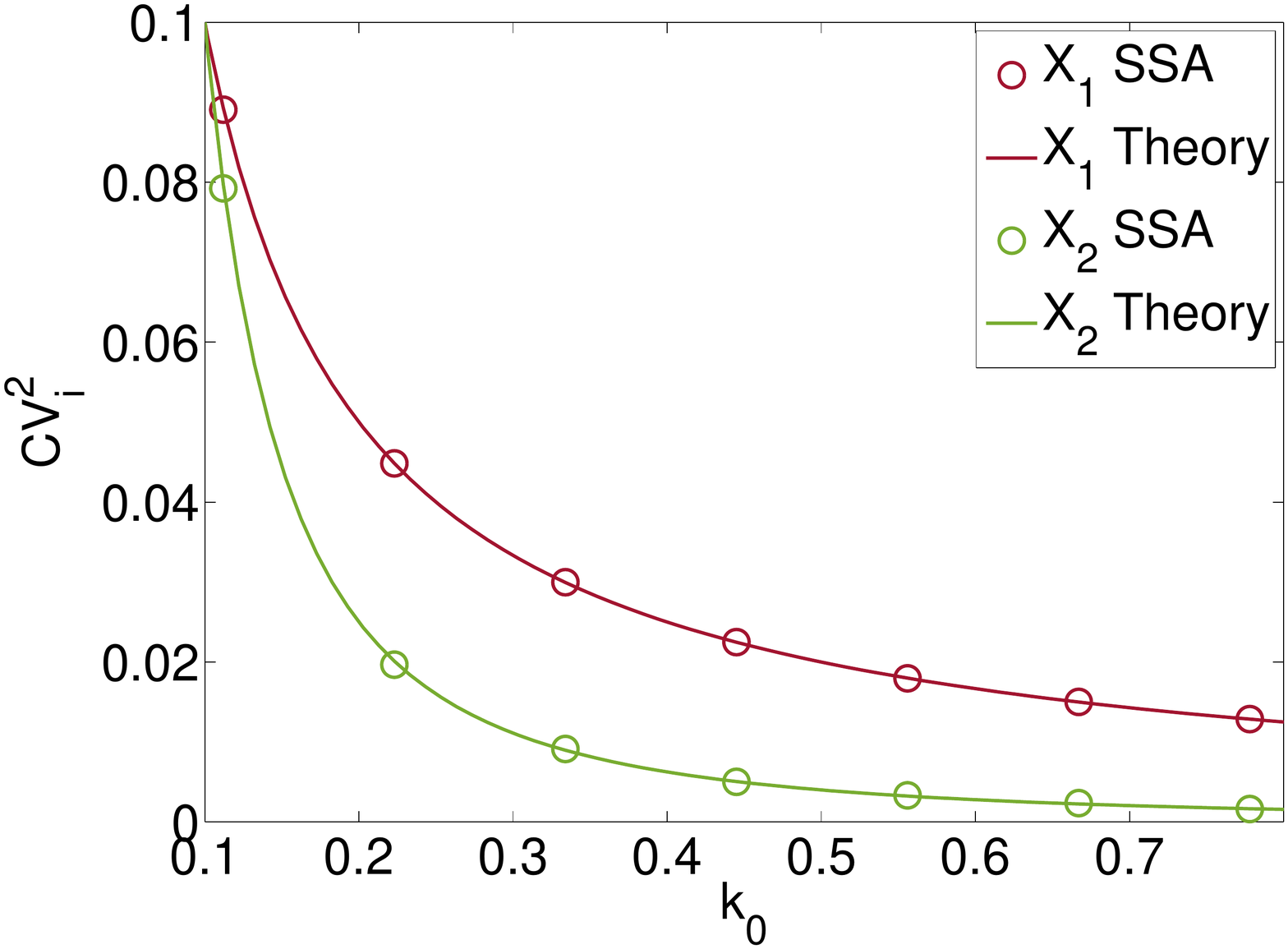}
}
\subfigure[]{
\includegraphics[width=50mm]{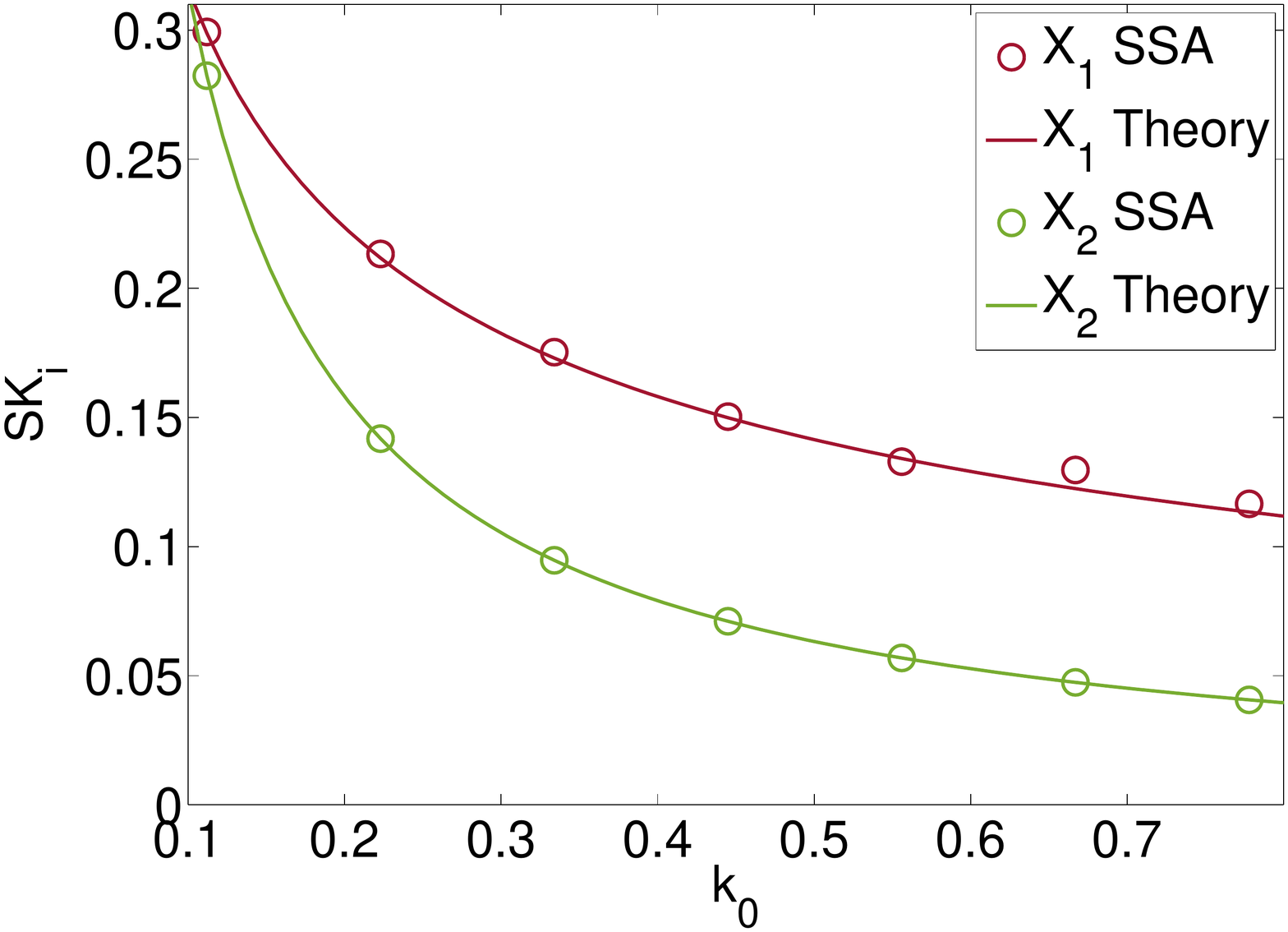}
}
\subfigure[]{
\includegraphics[width=50mm]{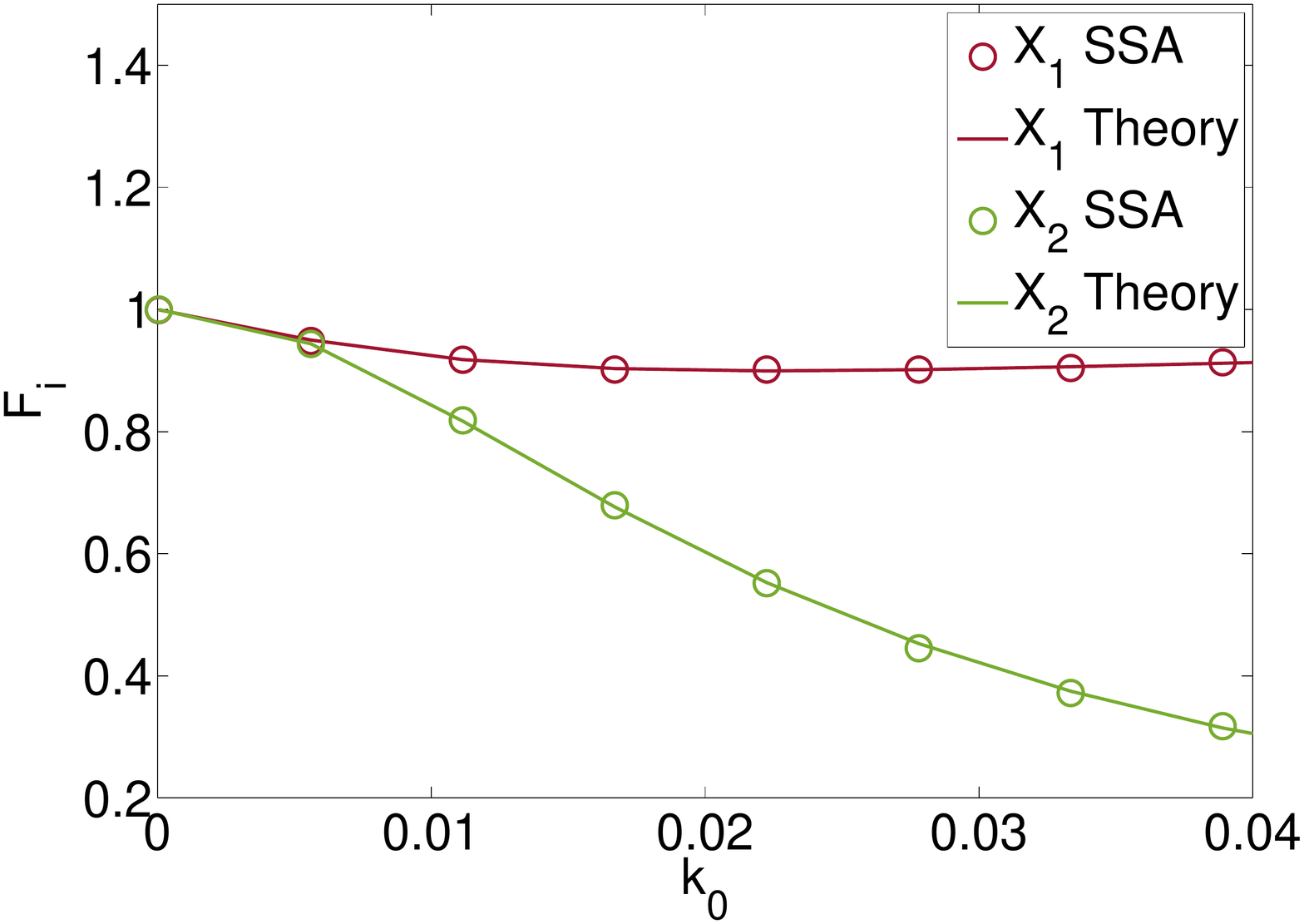}
}
\subfigure[]{
\includegraphics[width=50mm]{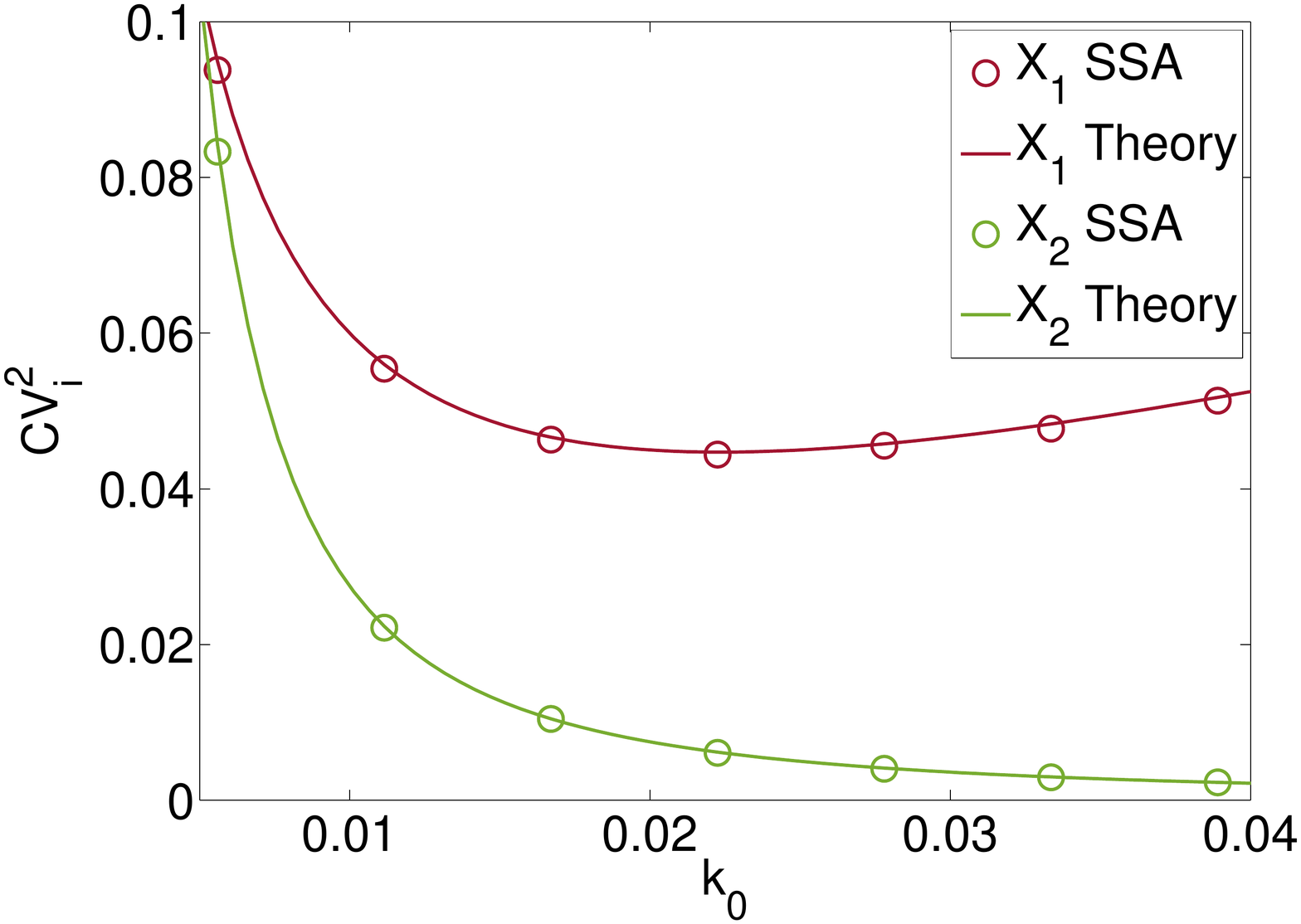}
}
\subfigure[]{
\includegraphics[width=50mm]{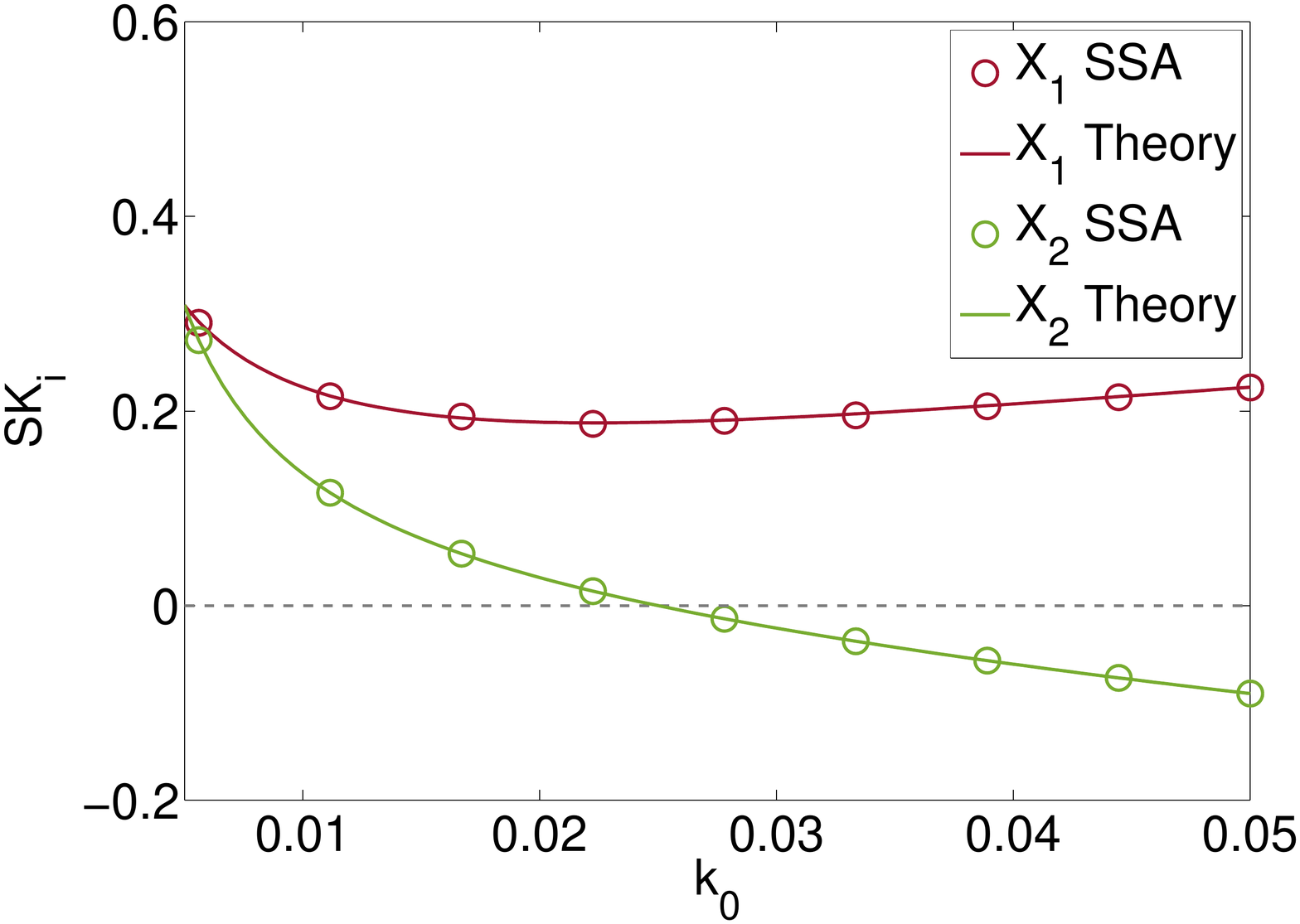}
}
\caption{Our theoretical predictions (lines) for the crowded and non-crowded models of the open homodimerisation reaction compared with data from the SSA of the CME and vCME (circles) for species $X_1$ (red) and $X_2$ (green). From left to right, we plot the Fano factor, coefficient of variation, and skewness as a function of the parameter $k_0$, using Eqs. (\ref{statsapp1a}) for the upper panel (dilute conditions, CME) and Eqs. (\ref{statsapp1b}) for the lower panel (volume exclusion conditions, vCME). The parameter values are: $k_1=k_2=k_3=0.1$, $\Omega=10,$ $N=200$.  We note that in the dilute limit $k_0\approx 0$, the two systems have the same behaviour.}
 \end{figure}

\section{Stochastic description of chemical systems with a special type of chemical conservation laws}

In this section we study systems with a chemical conservation law implying that the sum of the molecule numbers of some of the species is a constant $k$. For these systems we show that: \emph{(i) the marginal distribution of the global molecule numbers of a chemical species $X_i$ not involved in the conservation law is Poisson ($\Omega \phi_i$) if volume exclusion is ignored and Binomial ($N - k, \Omega \tilde{\phi}_i/(N-k)$) if it is taken into account. (ii) the marginal distribution of the global molecule numbers of a chemical species $X_i$ involved in the conservation law is Binomial ($k, \Omega \phi_i/k$) if volume exclusion is ignored and Binomial ($k, \Omega \tilde{\phi}_i/k$) if it is taken into account}. We shall call these Statement 2 and 3 respectively. We also discuss the physical implications of these statements, as well as confirm our results using stochastic simulations of the RDME and the vRDME applied to an open heterodimerisation reaction. 

\subsection{Derivation of Statements 2 and 3 and the dilute limit} 

We consider a chemical system with $M$ chemical species and a chemical conservation law of the form:
\begin{equation}\label{conslaw1}
 \sum_{i=L+1}^{M} n_i=k,
 \end{equation}
where $n_i$ is the number of molecules of species $i$, and $1\leq L \leq M-2$. This is a special case of the general global conservation law considered in Section III. It implies that there are no restrictions on the number of molecules of species $X_1,...,X_L$, but that the sum of the number of molecules of species $X_{L+1},...,X_M$ is constant at all times. 

The global probability distribution for $M$ chemical species, assuming point particles, Eq. (\ref{vKresult}), is then given by:
\begin{equation}
P(\vec{n})=k! e^{-\sum_{i=1}^L (\Omega\phi_i)} \frac{(\Omega\phi_{1})^{n_1}}{n_1!}...\frac{(\Omega\phi_{L})^{n_L}}{n_L!} \biggl(\frac{(\Omega\phi_{L+1}/k)^{n_{L+1}}}{n_{L+1}!}...\frac{(\Omega\phi_{M}/k)^{n_{M}}}{n_{M}!} \delta( \sum_{i=L+1}^{M} n_i,k) \biggr),
\end{equation}
which is the product of a multivariate Poisson distribution with parameters $(\{\Omega \phi_1,...,\Omega \phi_L\})$ and a multinomial distribution with parameters $(\{\Omega {\phi}_{L+1} / k,...,\Omega {\phi}_{M}/ k \},k)$. The multinomial originates from the constraint placed by the chemical conservation law Eq. (\ref{conslaw1}). Hence it follows, by the same arguments as in the previous section, that if a chemical species $X_i$ is not involved in the chemical conservation law, then the marginal distribution is Poisson $(\Omega \phi_i)$ whereas if it is involved in the chemical conservation law then the marginal distribution is Binomial $(k, \Omega \phi_i/k)$. 

The global probability distribution for $M$ chemical species, assuming a finite molecular radius, Eq. (\ref{vCMEsol}), specialised to the conservation law Eq. (\ref{conslaw1}), is given by:
\begin{align}
P(\vec{n})=&\frac{(N-k)!}{(N-k)^{N-k}} \biggl(\frac{(\Omega \tilde{\phi}_{1})^{n_{1}}}{n_{1}!}...\frac{(\Omega\tilde{\phi}_{L})^{n_{L}}}{n_{L}!} \frac{(\Omega\tilde{\phi}_{M+1})^{n_{M+1}}}{n_{M+1}!} \delta( \sum_{i=1}^{L} n_i + n_{M+1},N-k) \biggr) \nonumber \\ & \times k! \biggl(\frac{(\Omega\tilde{\phi}_{L+1}/k)^{n_{L+1}}}{n_{L+1}!}...\frac{(\Omega\tilde{\phi}_{M}/k)^{n_{M}}}{n_{M}!} \delta( \sum_{i=L+1}^{M} n_i,k) \biggr),
\end{align}
which is the product of a multinomial distribution with parameters $(\{\Omega \tilde{\phi}_{1} /(N-k),...,\Omega \tilde{\phi}_{L}/ (N-k), \Omega \tilde{\phi}_{M+1}/ (N-k) \},N-k)$ and a multinomial with parameters $(\{\Omega \tilde{\phi}_{L+1} / k,...,\Omega \tilde{\phi}_{M}/ k \},k)$. The latter multinomial originates from the chemical conservation law Eq. (\ref{conslaw1}). The former multinomial originates from the combination of the chemical conservation law Eq. (\ref{conslaw1}) and the volume exclusion law in the vRDME, $\sum_{i=1}^{M+1} n_i=N$, which together imply the combined conservation law $\sum_{i=1}^{L} n_i + n_{M+1}=N - k$. Hence it follows that if a chemical species $X_i$ is not involved in the chemical conservation law, then the marginal distribution is Binomial $(N-k, \Omega \tilde{\phi}_i/(N-k)$ whereas if it is involved in the chemical conservation law then the marginal distribution is Binomial $(k, \Omega \tilde{\phi}_i/k)$. It is straightforward to verify using the Poisson limit theorem that in the dilute limit the global Binomial solution of the vRDME approaches the Poisson solution of the RDME. 

\subsection{Statistical measures and Physical implications} 

For those species not involved in the chemical conservation law, the marginal is Poisson $(\Omega \phi_i)$ and hence the statistical quantities are given by Eq. (\ref{statsP}) if one assumes dilute conditions. If volume exclusion effects are considered then the marginal distribution is Binomial $(N-k, \Omega \phi_i/(N-k))$ and hence the statistics are given by Eq. (\ref{cqts}) with the parameter $N$ replaced by $N-k$. Similarly it can be reasoned that for those species involved in the chemical conservation law, i.e. species $X_{L+1},...,X_{M}$, the statistics are given by Eq. (\ref{cqts}) with the parameter $N$ replaced by $k$ and $\tilde \phi$ replaced by $\phi$ if dilute conditions are assumed and by Eq. (\ref{cqts}) with the parameter $N$ replaced by $k$ if volume exclusion is taken into account. 

The physical implication of these results is as follows. For both species which are involved and not involved in the chemical conservation law, taking into account volume exclusion implies that the marginal distribution is Binomial and hence we can make the same statement as for chemical systems without any chemical conservation laws. Namely for chemical systems with a chemical conservation law of the type Eq. (\ref{conslaw1}), {\emph{as the extent of volume exclusion increases, i.e, as $\langle n_i \rangle / N \rightarrow 1$, the fluctuations become more sub-Poissonian, deviations from the classical noise-strength power law become more apparent and the distribution of fluctuations changes from being skewed to the right (positive skewness) to being skewed to the left (negative skewness). The latter occurs when the $\langle n_i \rangle / (N - k)$ exceeds $1/2$ for species not involved in the chemical conservation law and when $\langle n_i \rangle / k$ exceeds $1/2$ for species involved in the chemical conservation law}}. 

However there are also some differences between the results here and those of the previous section. The RDME predicts the wrong qualitative dependence of the Fano factor, coefficient of variation and the skewness on the mean molecule numbers for all species in the system without any chemical conservation law. For systems with a chemical conservation law, this is still true for those species not involved in a chemical conservation law. However the RDME does predict the right qualitative dependence for those species involved in the chemical conservation law (since it predicts a Binomial marginal distribution, same as the vRDME, though with different parametrisation). 

The results here can also be generalised to a system with a number of chemical conservation laws of the sum type. For example for a system with two conservation laws of the type:
\begin{equation}\label{conslaw1n}
 \sum_{i=z+1}^{L} n_i=s, \sum_{i=L+1}^{M} n_i=k,
 \end{equation}
where $1\leq z \leq L-2$, by a similar reasoning as above, we find, assuming a finite molecular radius, that the marginal distributions of species $X_i$ is Binomial $(N-k-s, \Omega \tilde{\phi}_i/(N-k-s))$ if it is not involved in the conservation laws, is Binomial $(s, \Omega \tilde{\phi}_i/s)$ if it is involved in the first conservation law and Binomial $(k, \Omega \tilde{\phi}_i/k)$ if it is involved in the second conservation law.

\subsection{Application: open heterodimerisation reaction}

We now consider the dilute (point particle) model of the chemical system:
\begin{align}\label{system7}
\emptyset  \xrightleftharpoons[k_1]{k_0} X_1,\quad
X_1+X_2 \xrightleftharpoons[k_3]{k_2}X_3,
\end{align}
whereby a species $X_1$ is produced and subsequently molecules of this species and that of $X_2$ reversibly bind to form molecules of $X_3$, a heterodimer. The system has the implicit chemical conservation law $n_2 + n_3 = k$ where $k$ is a time-independent constant determined by the initial conditions, and hence it is of the type studied above. The deterministic equilibrium constants are:
\begin{equation}
\label{s7eqs}
\phi_1 = \frac{k_0}{k_1}, \quad \frac{\phi_1 \phi_2}{\phi_3} = \frac{k_3}{k_2}.
\end{equation}

The volume excluded version (assuming finite molecular size) of reaction scheme (\ref{system7}) is:
\begin{align}\label{system8}
X_4  \xrightleftharpoons[\tilde{k}_1]{\tilde{k}_0} X_1,\quad
X_1+X_2 \xrightleftharpoons[\tilde{k}_3]{\tilde{k}_2}X_3 + X_4,
\end{align}
where $X_4$ is the empty space species and now we have two conservation laws: the chemical law $n_2 + n_3 = k$ and the volume exclusion law $n_1+n_2+n_3+n_4 = N$ where $N$ is the maximum number of molecules which the compartment can accommodate. The deterministic equilibrium constants are:
\begin{equation}
\label{s8eqs}
\frac{\tilde{\phi}_1}{\tilde{\phi}_4} = \frac{\tilde{k}_0}{\tilde{k}_1} = \frac{k_0 \Omega}{k_1 N}, \quad \frac{\tilde{\phi}_1 \tilde{\phi}_2}{\tilde{\phi}_3 \tilde{\phi}_4} = \frac{\tilde{k}_3}{\tilde{k}_2} = \frac{k_3 \Omega}{k_2 N},
\end{equation}
where we used the relationship between the rate constants of the volume excluded and dilute systems (as in the previous example and as elucidated in Section IV). 

Explicit solution of Eqs. (\ref{s7eqs}) and Eqs. (\ref{s8eqs}) together with the relevant conservation laws leads to:
\begin{equation}
\frac{\phi_1}{\tilde{\phi}_1} = 1 + \frac{k k_1 + k_0 \Omega}{k_1 (N - k)}, \quad \phi_2 = \tilde{\phi}_2 = \frac{k k_1 k_3}{(k_0 k_2 + k_1 k_3) \Omega}, \quad \phi_3 = \tilde{\phi}_3= \frac{k k_0 k_2}{(k_0 k_2 + k_1 k_3) \Omega}.
\end{equation}
This implies that the concentrations of species $X_2$ and $X_3$ (the species involved in a chemical conservation law) are insensitive to volume exclusion effects but the concentration of species $X_1$ is found to decrease when crowding is taken into account. Intuitively this because species $X_2$ and $X_3$ are involved in a chemical conservation law and hence the impact of the second conservation law due to volume exclusion is nullified; species $X_1$ in contrast is not involved in any chemical conservation law and is hence strongly affected by the conservation law due to volume exclusion. 

According to our theory in the previous section, (i) the marginal global distribution of species $X_1$ is Poisson $(\Omega \phi_1)$ according to the RDME and Binomial $(N-k,\Omega \tilde{\phi}_1/(N-k))$ according to the vRDME. The mean of the latter is less than that of the former. This is verified via stochastic simulations of the RDME and vRDME using the SSA -- see Fig. 4(a); (ii) the marginal global distribution of species $X_2$ is Binomial $(k,\Omega \phi_2/k))$ for both the RDME and vRDME. This is also verified via stochastic simulations using the SSA -- see Fig. 4(b). In Fig. 4(c) and 4(d) we also show that stochastic simulations using the SSA agree with the theoretical expressions obtained by marginalising the local (voxel) distributions given by Eqs. (\ref{RDMElocal}) and (\ref{vRDMElocal}) in Section III. Note that for the purpose of stochastic simulations using the RDME and vRDME, we need to specify a lattice type. We choose the RDME and vRDME lattices to be periodic, square and in two dimensions, with the neighbourhood of a voxel being the four cells orthogonally surrounding it. The compartment volume $\Omega$ will be fixed to one, meaning that for $N$ voxels, the lattice consists of $\sqrt{N} \times \sqrt{N}$ voxels with lattice spacing $1/\sqrt{N}$. We shall use this lattice for all stochastic simulations in this article.

\begin{figure} [h]
\centering
\subfigure[\ Global distribution for species $X_1$]{
\includegraphics[width=65mm]{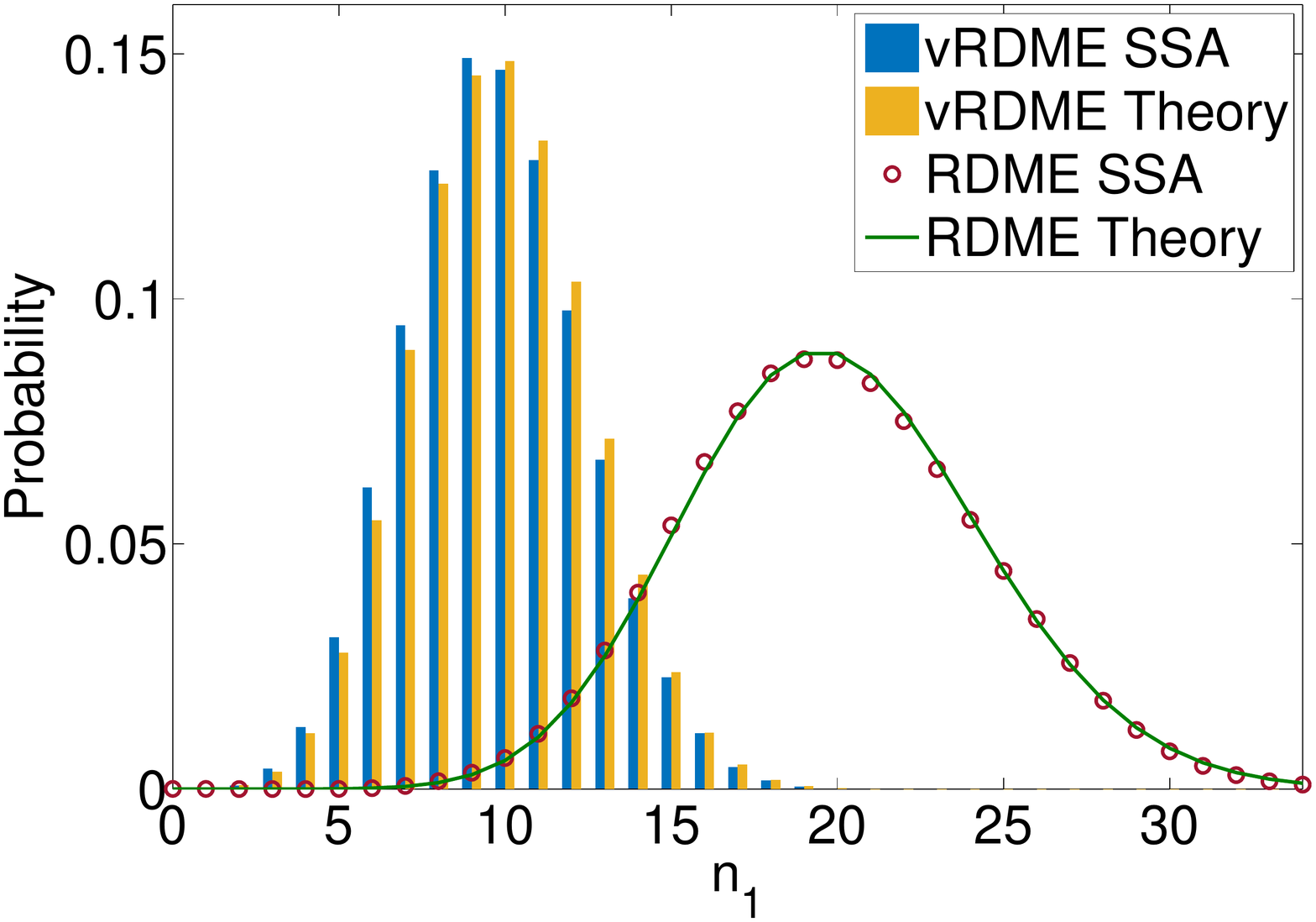}
\label{fig:subfig1}
}
\subfigure[\ Global distribution for species $X_2$]{
\includegraphics[width=65mm]{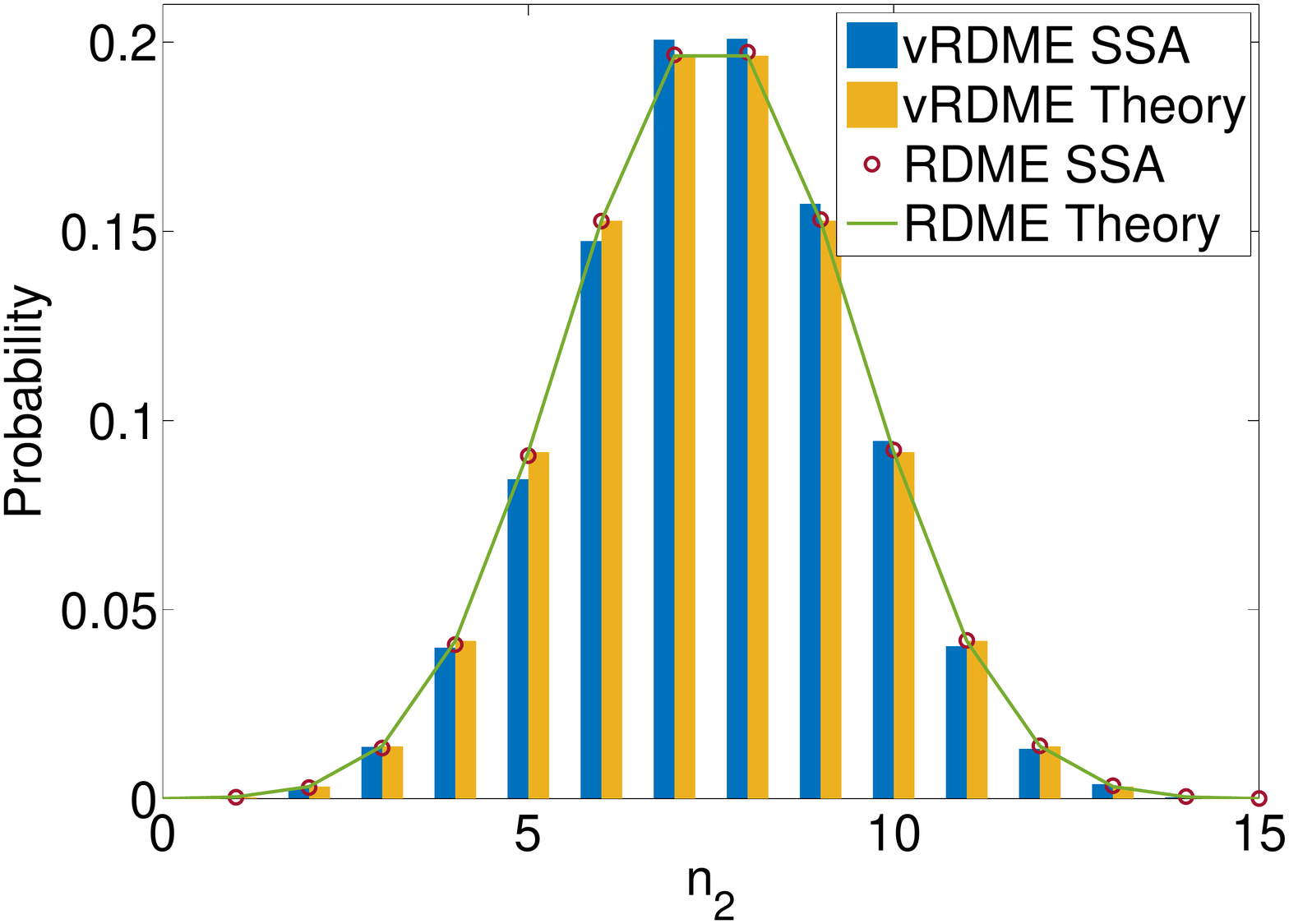}
\label{fig:subfig2}
}
\subfigure[\ Local (voxel) distribution for $X_1$]{
\includegraphics[width=65mm]{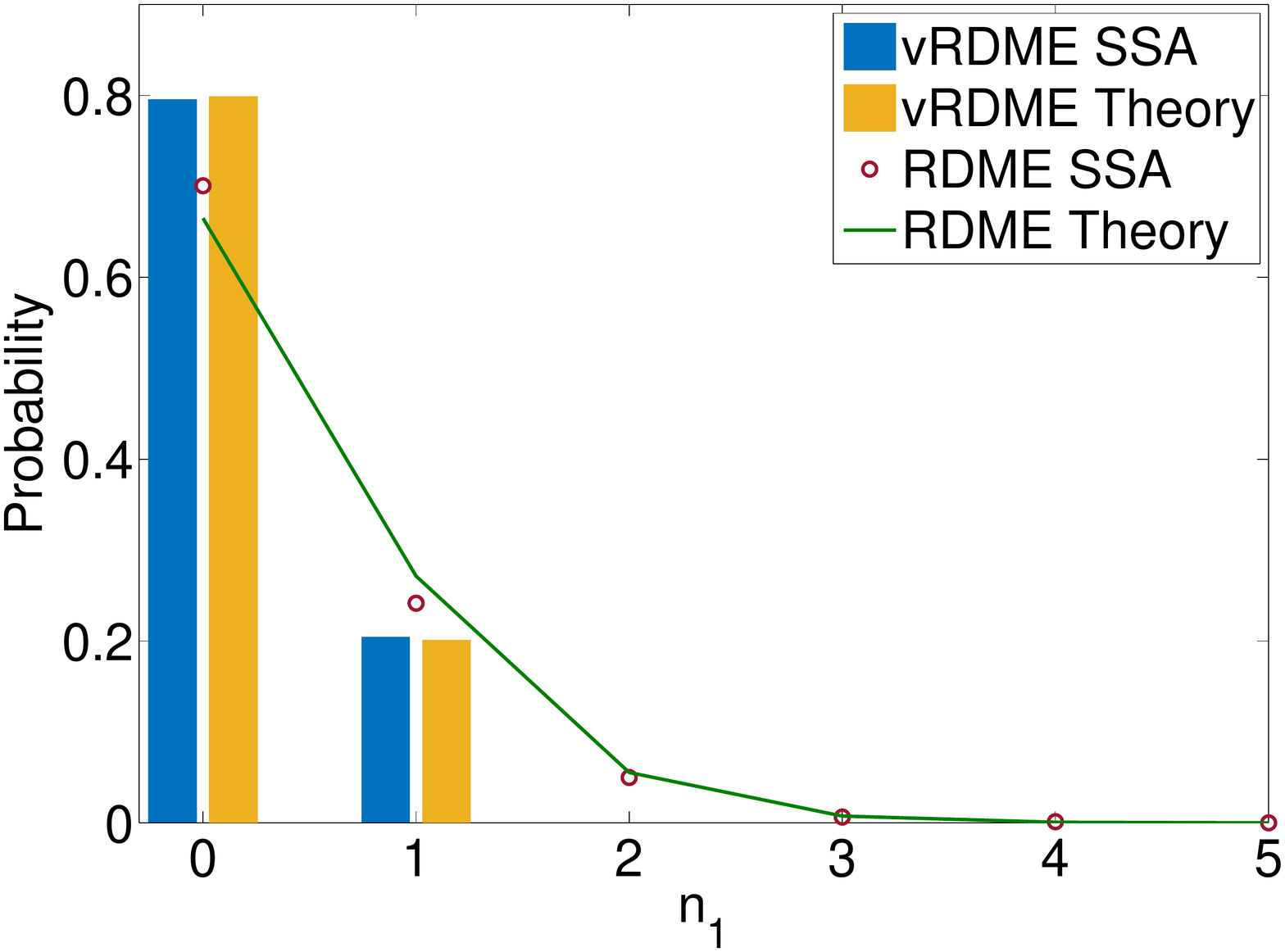}
\label{fig:subfig1}
}
\subfigure[\ Local (voxel) distribution for $X_2$]{
\includegraphics[width=65mm]{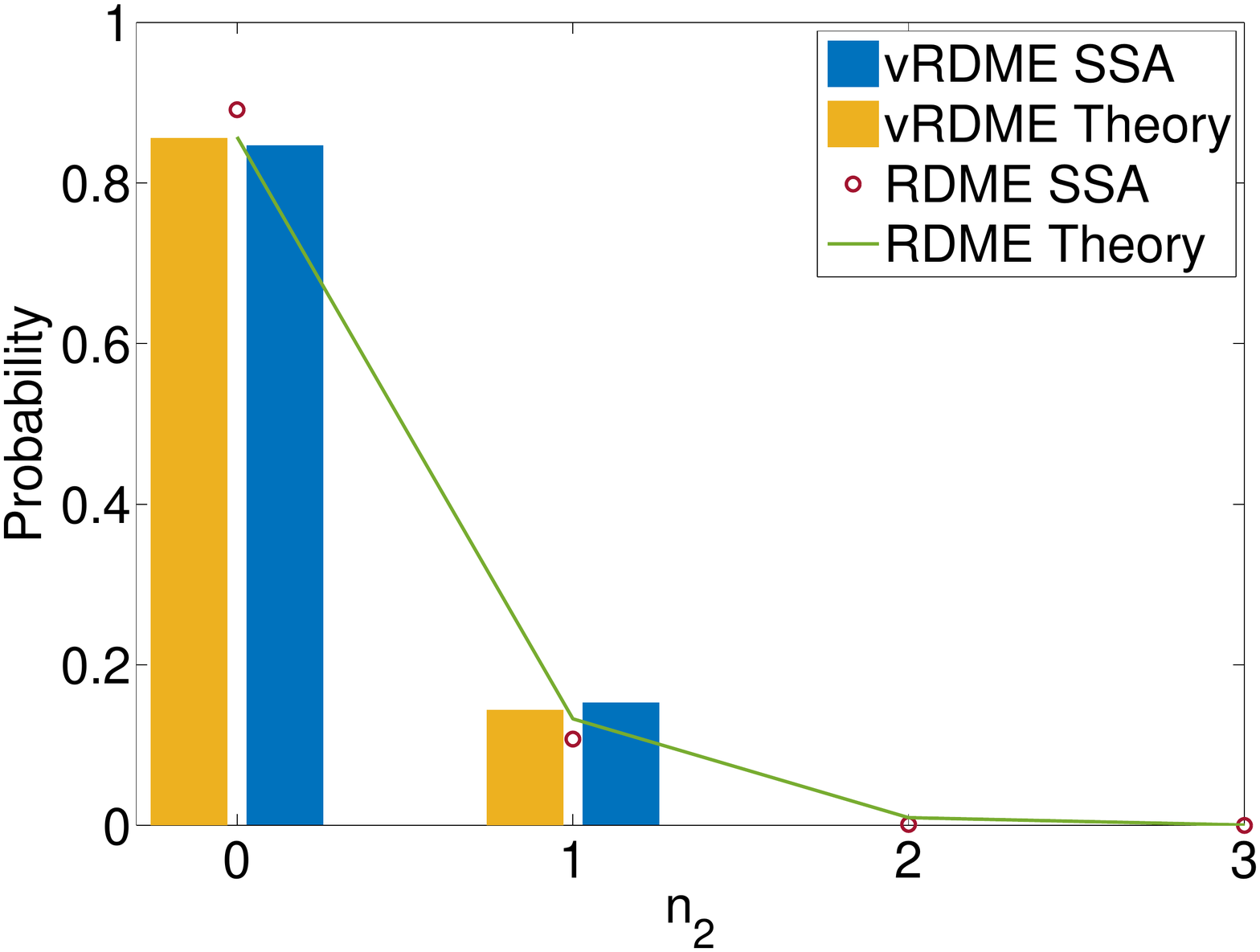}
\label{fig:subfig2}
}
\caption{Comparisons of vRDME and RDME simulations with our theoretical predictions for the local and global distributions of molecule numbers of species $X_1$ and $X_2$ in the heterodimerisation system. Parameter values are $k_0=20$, $k_1=1$, $k_2=1$ and $k_3=20$ and the chemical conservation law is $n_2 + n_3 = k =15$. The global compartment volume is $\Omega=1$ and the total number of voxels for both the RDME and vRDME is $N=49$. In all cases there is excellent agreement between simulations and theory.}
 \end{figure}

Of interest is how the vRDME probability distribution of the global number of molecules of species $X_1$ changes as the ratio of molecular diameter to compartment length scale is varied. The ratio of the compartment side length to the molecular diameter (the lattice spacing) is given by $\sqrt{N}$. In Fig. 5 (a) we plot the global marginal probability distribution solution of the vRDME for species $X_1$, i.e., Binomial $(N-k,\Omega \tilde{\phi}_1/(N-k))$, as a function of the total number of voxels $N$ while keeping the compartment volume constant. Good agreement of the vRDME and RDME solutions is obtained when $N = 1600$, i.e, when the compartment side length is about forty times larger than the molecular diameter; here the molecules are small enough that the system is dilute. In contrast, clear differences exist between the vRDME and RDME predictions when $N = 100$ (and smaller values) which corresponds to the case of molecules whose diameter is at least $1/10$ of the compartment side length; for these cases the RDME overpredicts the true global concentration of $X_1$. 

It is also interesting to understand the effects of increasing the degree of volume exclusion by adding inert molecules to the chemical reaction system. This is of particular relevance to understanding intracellular reaction systems which typically operate in such conditions, i.e., molecules of other intracellular pathways which are inert with respect to the reaction system of interest exert influence on the latter via volume exclusion effects \cite{ZimmermannMinton1993}. To this end we consider a modified version of reaction scheme (\ref{system8}):
\begin{align}\label{system9}
X_4  \xrightleftharpoons[\tilde{k}_1]{\tilde{k}_0} X_1,\quad X_1+X_2 \xrightleftharpoons[\tilde{k}_3]{\tilde{k}_2}X_3 + X_4, \quad X_4  \xrightleftharpoons[\tilde{k}_5]{\tilde{k}_4} X_5,
\end{align}
where $X_4$ is the empty space species and $X_5$ is a chemical species which does not chemically interact with the rest of the molecules (an inert external crowder). In Fig. 5(b) we plot the global marginal probability distribution solution of the vRDME for species $X_1$, i.e., Binomial $(N-k,\Omega \tilde{\phi}_1/(N-k))$, as a function of the mean number of inert external crowder molecules $\langle n_5 \rangle$. Note that the effect of increasing molecular crowding by adding more molecules of $X_5$ is to induce a shift of the probability distribution to the left such that there are fewer molecules, on average, of $X_1$ in the system. This is qualitatively similar to the effect seen in Fig. 5(a). This similarity arises because an increase in the fraction of occupied space can either be induced by increasing the size of the reactant molecules while keeping the compartment size fixed (the case of Fig. 5(a)) or else by introducing inert molecules into the system (the case of Fig. 5(b)). Note that in both cases the marginal distribution of $X_2$ is unchanged by the degree of volume exclusion since as we noted earlier both the RDME and vRDME give the same result. 

\begin{figure} [h]
\centering
\subfigure[\ Global distribution for species $X_1$ as a function of reactant molecule size]{
\includegraphics[width=75mm]{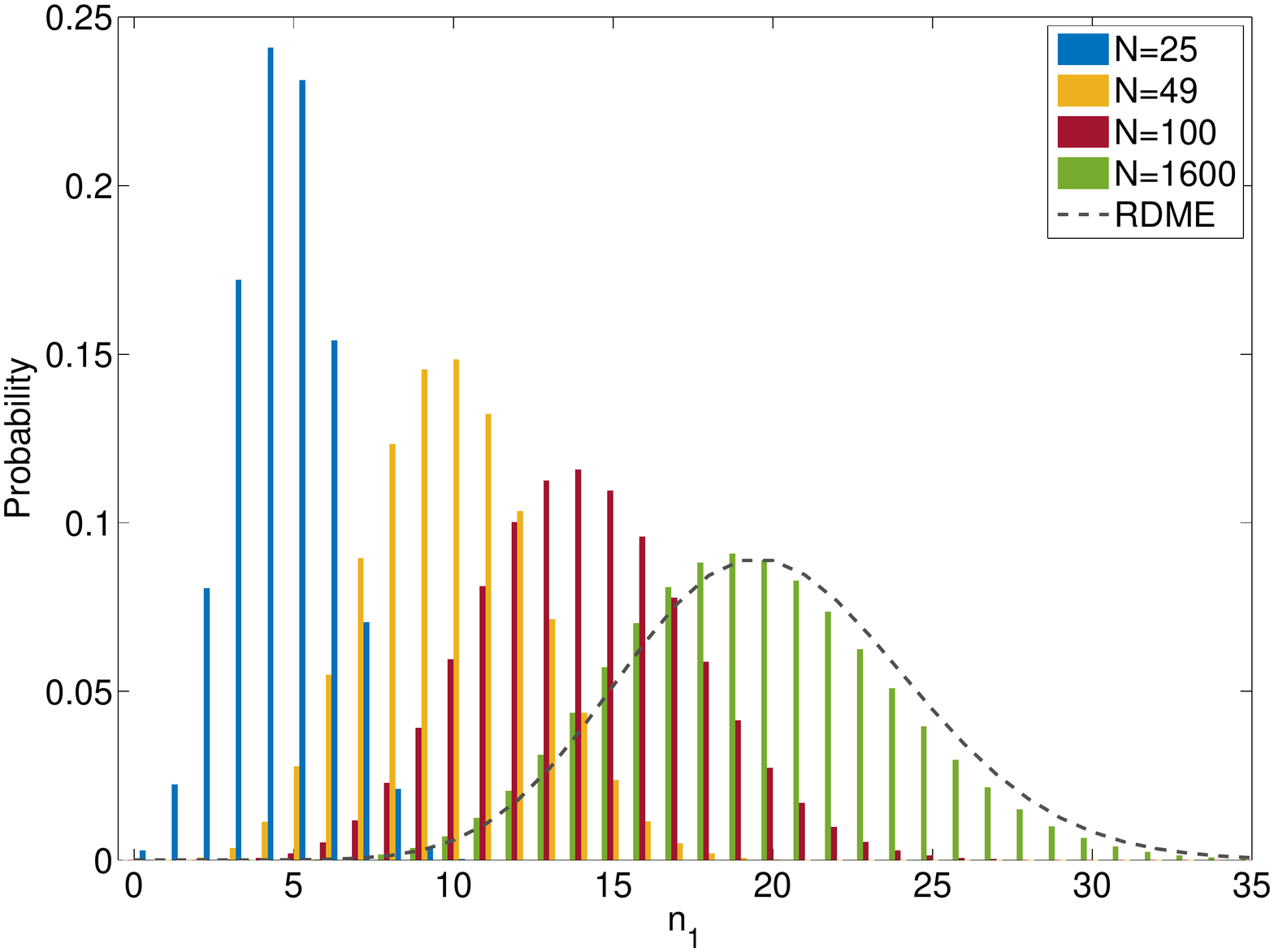}
}
\subfigure[\ Global distribution for species $X_1$ as a function of mean \# of inert crowders]{
\includegraphics[width=75mm]{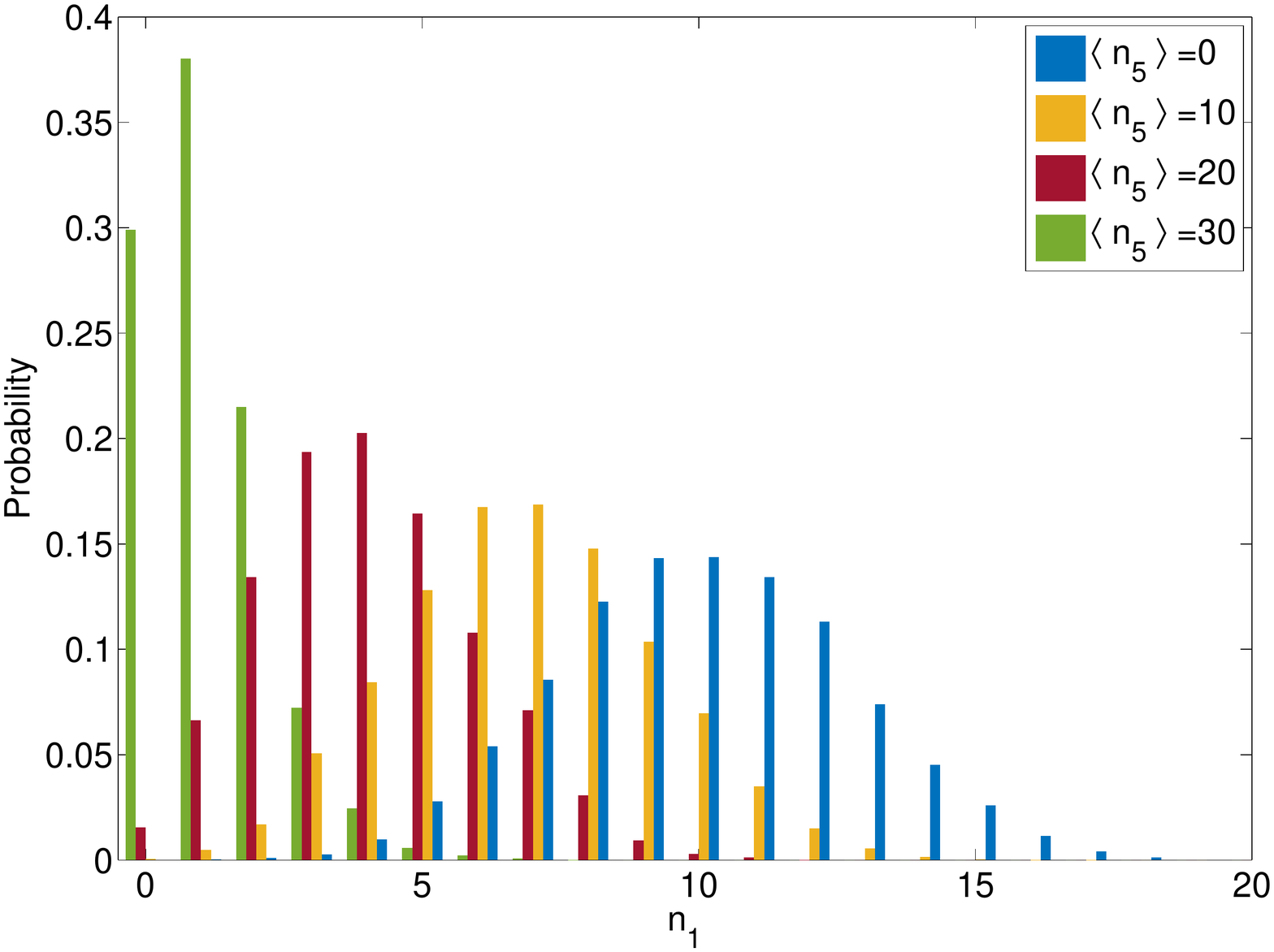}
}
\caption{Variation of the vRDME global distribution of molecule numbers for species $X_1$ in the open heterodimerisation reaction, as a function of the occupied volume fraction of space. The degree of volume exclusion is controlled by varying the size of the reactant molecules in (a) and by introducing inert molecules into the system in (b). Specifically (a) is obtained by keeping the compartment size constant and varying the maximum number of molecules $N$ (voxels) which can be accommodated in the compartment for system (\ref{system8}). While (b) is obtained by varying the ratio $\tilde{k}_4/\tilde{k}_5$ which controls the mean number of inert external crowders $\langle n_5 \rangle$ in system (\ref{system9}). The parameters $k,k_0,k_1,k_2,k_3,\Omega$ for both (a) and (b) are the same as in Fig. 5. See text for discussion.}
 \end{figure}

\section{Stochastic description of chemical systems with more general chemical conservation laws}

Previously we have considered chemical conservation laws of the type (\ref{conslaw1}). Though common, these are not the only chemical conservation laws in nature. The general theory presented in Section III also applies to these other conservation laws. In what follows we use the latter results to study an example of a chemical system constrained by a chemical conservation law which is not of the sum type. In particular, we will show that in this case, the global marginal distribution of the number of molecules for a species involved in the conservation law is not Binomial, unlike the case of a species involved in a chemical conservation law of the type (\ref{conslaw1}).

\subsection{Closed dimerisation reaction}\label{supois}

Consider the point particle model of the reaction system:
\begin{equation}
\label{closeddimr}
X_1+X_1\xrightleftharpoons[k_1]{k_0} X_2,
\end{equation}
whereby two molecules of $X_1$ reversibly bind to form a dimer $X_2$. This system has the global chemical conservation law $n_1 + 2n_2=k$ where $k$ is a time-independent constant fixed by the initial conditions and hence it is not of the same type as the chemical conservation laws (\ref{conslaw1}) considered earlier. According to Eq. (\ref{vKresult}) and Eq. (\ref{vCMEsol}) the (normalised) marginal probability distribution solution for species $X_2$ according to the CME and the vCME is given by:
\begin{align}
\label{CMEcd}
P(n_2) &= \frac{2^{-k} (-1)^{k/2} (\frac{k_0}{k_1 \Omega})^{n_2 - k/2}k! }{(k - 2 n_2)! n_2! H_U[-\frac{k}{2},\frac{1}{2},-\frac{k_1 \Omega}{4 k_0}]}, \\ P(n_2) &= \frac{\Gamma(1+k) (\frac{k_0 N}{k_1 \Omega})^{n_2}}{(k - 2n_2)! n_2! (N + n_2 - k)! H_{2F1}^R [\frac{1}{2},-\frac{k}{2},-\frac{k}{2},1-k+N,\frac{4k_0 N}{k_1 \Omega}]}, \label{vCMEcd}
\end{align}
respectively. Here we have used the notation $H_U$ and $H_{2F1}^R$ to denote Tricomi's confluent hypergeometric function and the regularised hypergeometric function respectively (these are the functions HypergeometricU and Hypergeometric2F1Regularised in Mathematica's notation \cite{Mathematica}). Note that for the vCME, we have here considered the volume excluded version of reaction scheme (\ref{closeddimr}), namely $X_1+X_1\xrightleftharpoons[\tilde{k}_1]{\tilde{k}_0} X_2 + X_3$ with $X_3$ representing the empty space species and the equilibrium constant $\frac{\tilde{k}_0}{\tilde{k}_1} = \frac{k_0 N}{k_1 \Omega}$ (as elucidated in Section IV). All the statistics of the molecule numbers of species $X_1$ can be deduced from those of $X_2$ given the  conservation law $n_1 + 2n_2=k$. 

There are here clearly differences from what we previously found for chemical species involved in chemical conservation laws of the type (\ref{conslaw1}). While for the latter, the global marginal distributions where binomial independent of whether volume exclusion is taken into account or not (see Section VI), in the example presented in this section, the global marginal distributions are not binomial. This difference can be traced to the fact that a binomial originates as the marginal of a multinomial distribution and the latter is effectively a product of Poissons constrained by a rule stating that the sum of the fluctuating variables is a constant; this rule is naturally obeyed by systems in which the chemical conservation law is of the type (\ref{conslaw1}). 

In Fig. 6 we plot the steady-state probability distribution of global molecule numbers according to the CME and vCME for the case when $N = k$, i.e., the minimum number of voxels required to accommodate the maximum number of molecules allowed by the dimerisation reaction. We note that while the chemical conservation law shielded the effects of volume exclusion law for those species involved in laws of the type (\ref{conslaw1}) (see Fig. 4b), no such shielding occurred in the example here, as can be appreciated from the large difference between the two distributions in Fig. 6. Likely, the implicit mathematical reason for these differences is that for systems in Section VI, the chemical conservation law $\sum_{i=L+1}^{M} n_i = k$ is ``nested'' within the volume exclusion law $\sum_{i=1}^{M+1} n_i = N$, while no such nesting is present in the current example where the chemical conservation law is $n_1 + 2n_2=k$. 

\begin{figure} [h]
\includegraphics[width=75mm]{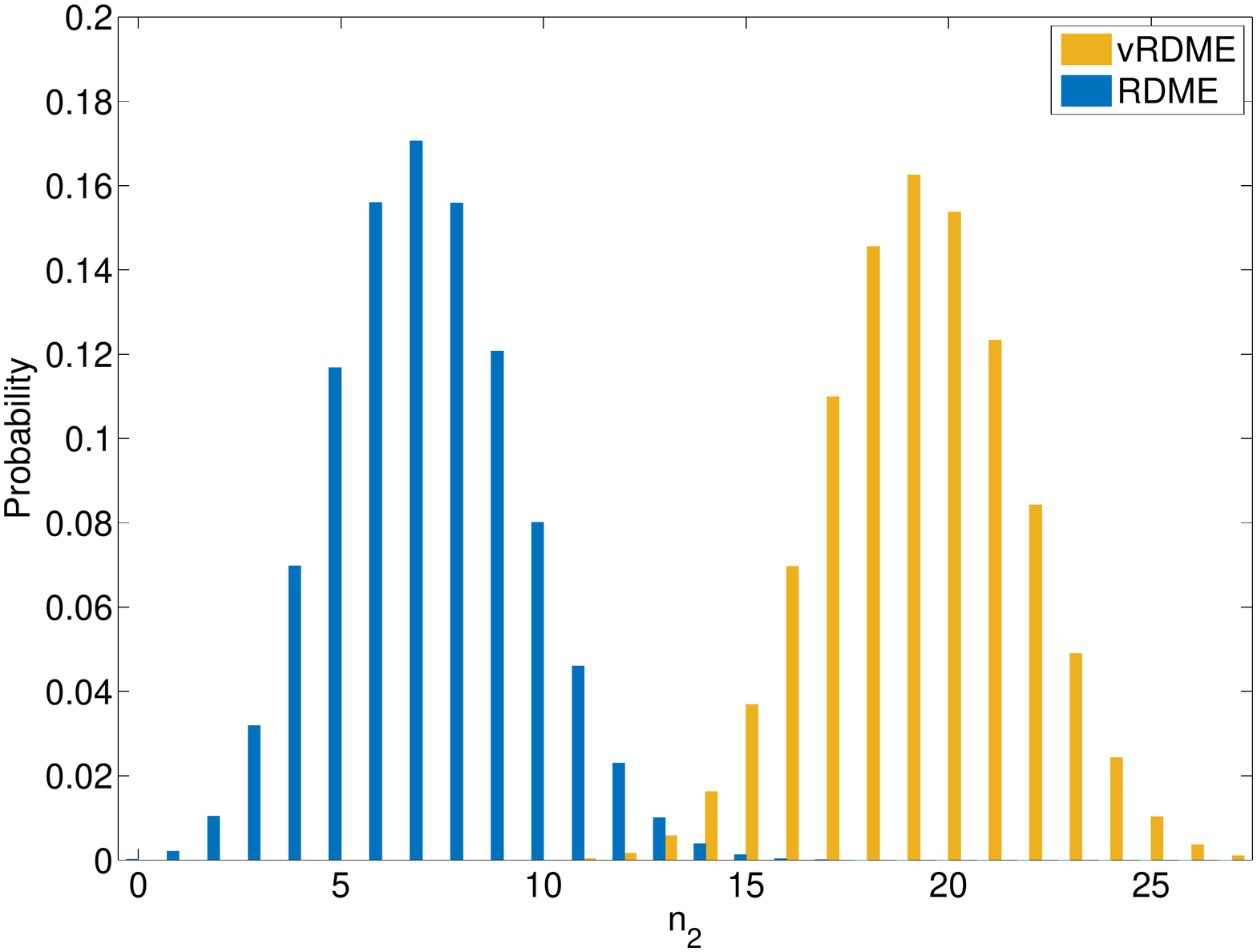}
\caption{The steady-state probability distribution according to the CME (dilute conditions) and to the vCME (volume exclusion is taken into account). The plots are generated using Eqs. (\ref{CMEcd})-(\ref{vCMEcd}). The distribution is significantly shifted by excluded volume effects; this is for the case where the maximum number of molecules which can be put inside a compartment of unit volume is $N = 100$. The reaction rate constants are $k_0 = 0.001$ and $k_1 = 1$, while the conservation law constant is $k = 100$.}
 \end{figure}

We now study the effect of volume exclusion on various statistics. We first note that the first and second moments of the vCME solution can be conveniently written in terms of three non-dimensional parameters: $k$, $R=N/k$ and $L=4 k_0/(k_1 \Omega)$. The parameter $L$ contains information about all the rate constants of the system; it is proportional to the equilibrium constant $k_0/k_1$ of the reaction in the absence of volume exclusion. The parameter $R$ is an inverse measure of volume exclusion. This is since as $N$ increases at constant compartment volume $\Omega$, molecules ``become smaller'' and hence the system becomes more dilute. The maximum degree of volume exclusion occurs when $N = k$, i.e, $R = 1$ and the dilute limit occurs when $R \rightarrow \infty$. In Fig. 7 we fix $k = 50$ and use Eq. (\ref{vCMEcd}) to calculate the statistics in very dilute conditions ($R = 1000$) and in highly crowded conditions ($R = 1$) as a function of the parameter $L$. The dilute statistics agree very well with those which can be calculated directly from the CME using Eq. (\ref{CMEcd}). 

In particular we find that: (i) the Fano Factor of species $X_2$ is always less than one and hence the distribution is sub-Poissonian in both volume excluded and dilute conditions (see Fig. 7a); (ii) the Fano factor of species $X_1$ can be greater than or less than 1 leading to three distinctive phases: sub-Poisson statistics in both volume excluded and dilute conditions (for $L < 7$), super-Poissonian in both conditions (for $L > 11$) and lastly a phase in which volume exclusion leads to a change from sub-Poissonian to super-Poissonian statistics (for $7 \le L \le 11$, see Fig. 7b). This is in contradistinction to the results in Section VI where we found that a species involved in chemical conservation laws of the type (\ref{conslaw1}) has sub-Poissonian fluctuations in both volume excluded and dilute conditions; (iii) volume exclusion leads to a decrease in the coefficient of variation of species $X_2$ and to an increase in the coefficient of variation of species $X_1$ (see Fig. 7c); (iv) volume exclusion leads to an increase in the mean number of molecules of species $X_2$ and to a decrease in the number of molecules of species $X_1$ (see Fig. 7d). Thus the inclusion of volume exclusion causes a shift of the equilibrium to the right, namely it leads to the production of more $X_2$ molecules and of less $X_1$ molecules. This is in agreement with the standard thermodynamic theory by Minton and co-workers \cite{Zhou2008}. We have numerically verified that these results hold for even $k$. 

As we saw in this example, the general properties of systems with chemical conservation laws of a general type are not typically open to analytical investigation because of the complicated form of the exact steady-state probability distribution solution of the CME and vCME. Nevertheless these expressions can be easily investigated numerically which is advantageous compared to lengthy stochastic simulations. 

\begin{figure} [h]
\centering
\subfigure[\ ]{
\includegraphics[width=65mm]{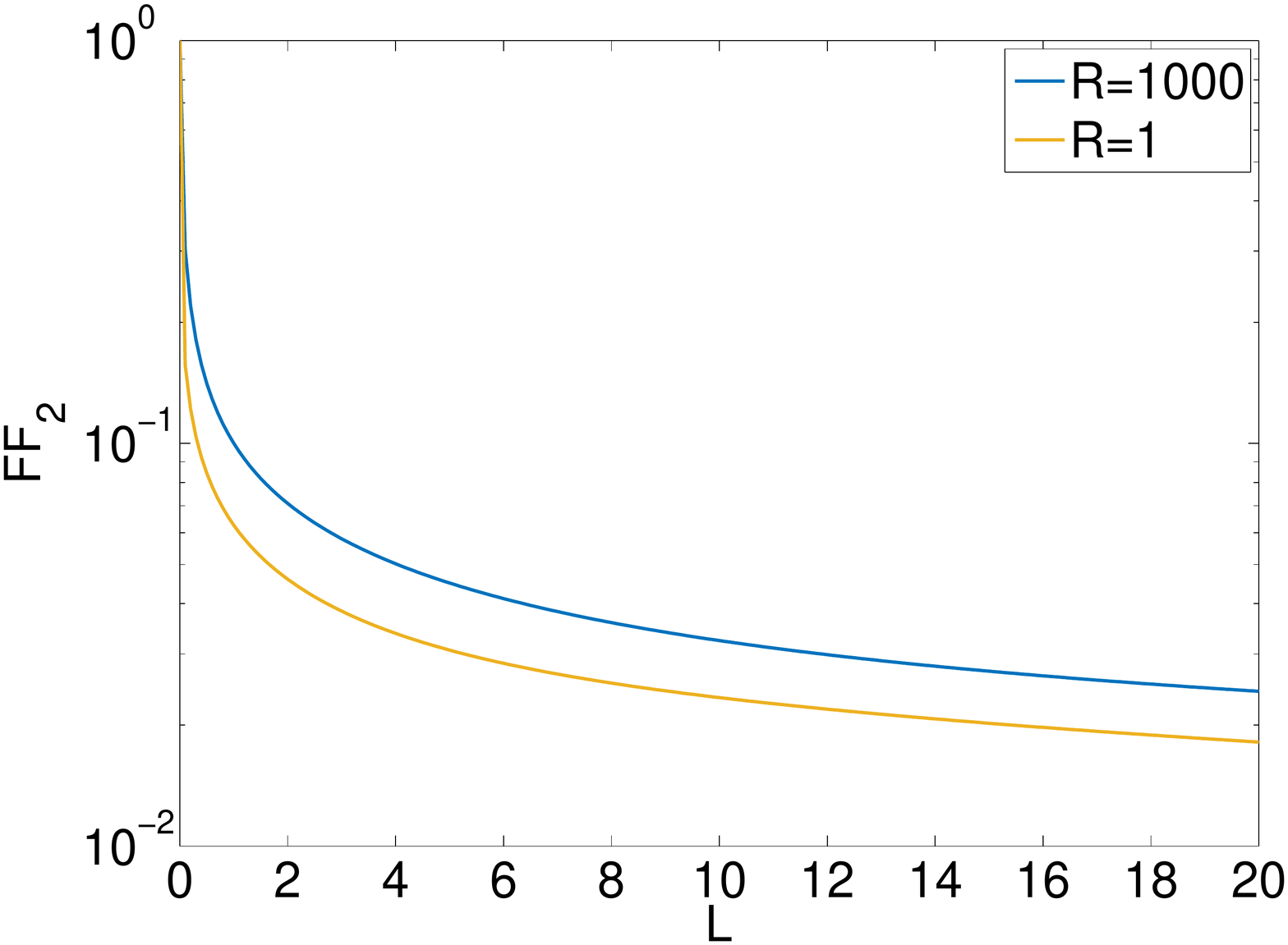}
}
\subfigure[\ ]{
\includegraphics[width=65mm]{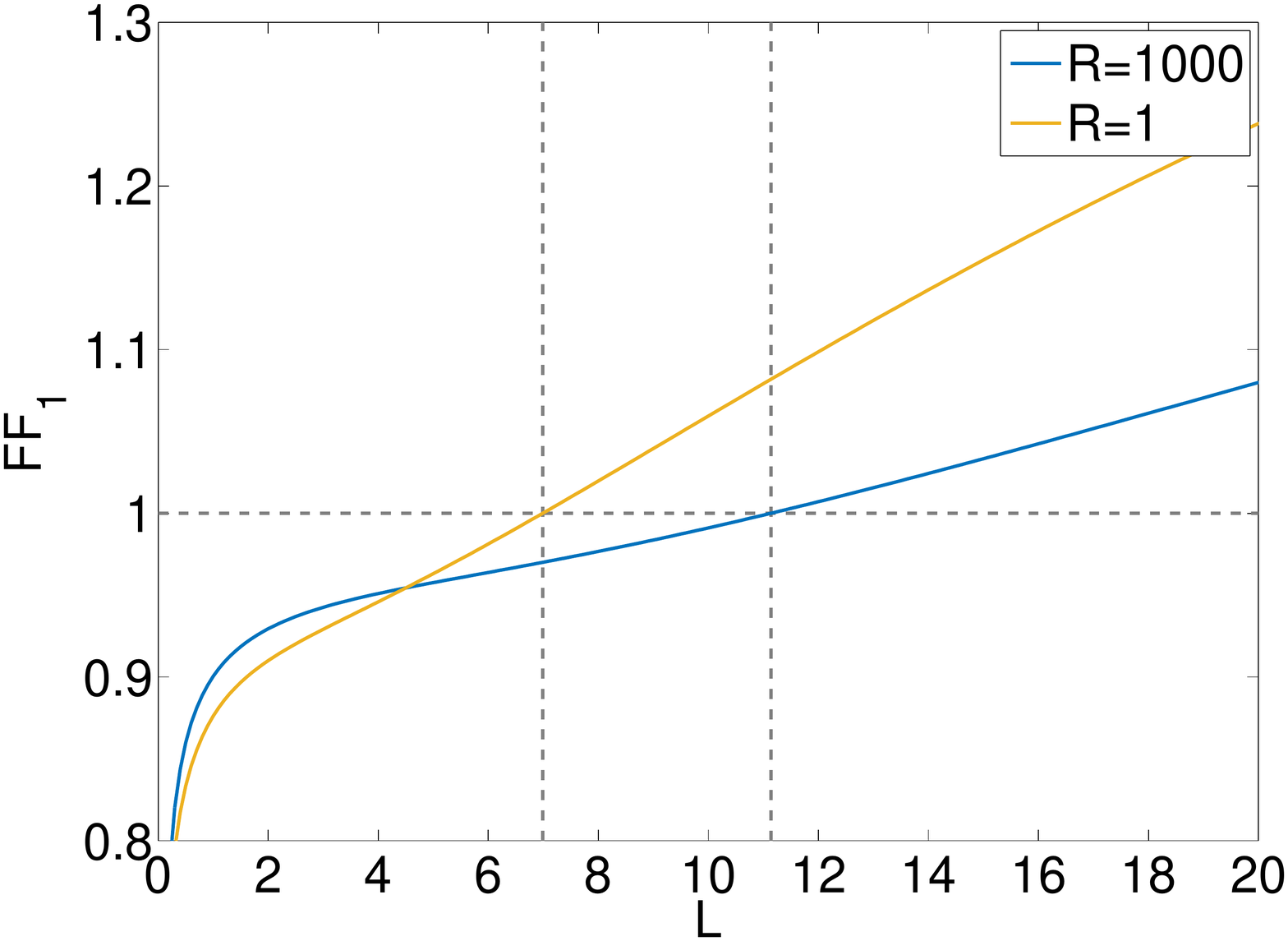}
}
\subfigure[\ ]{
\includegraphics[width=65mm]{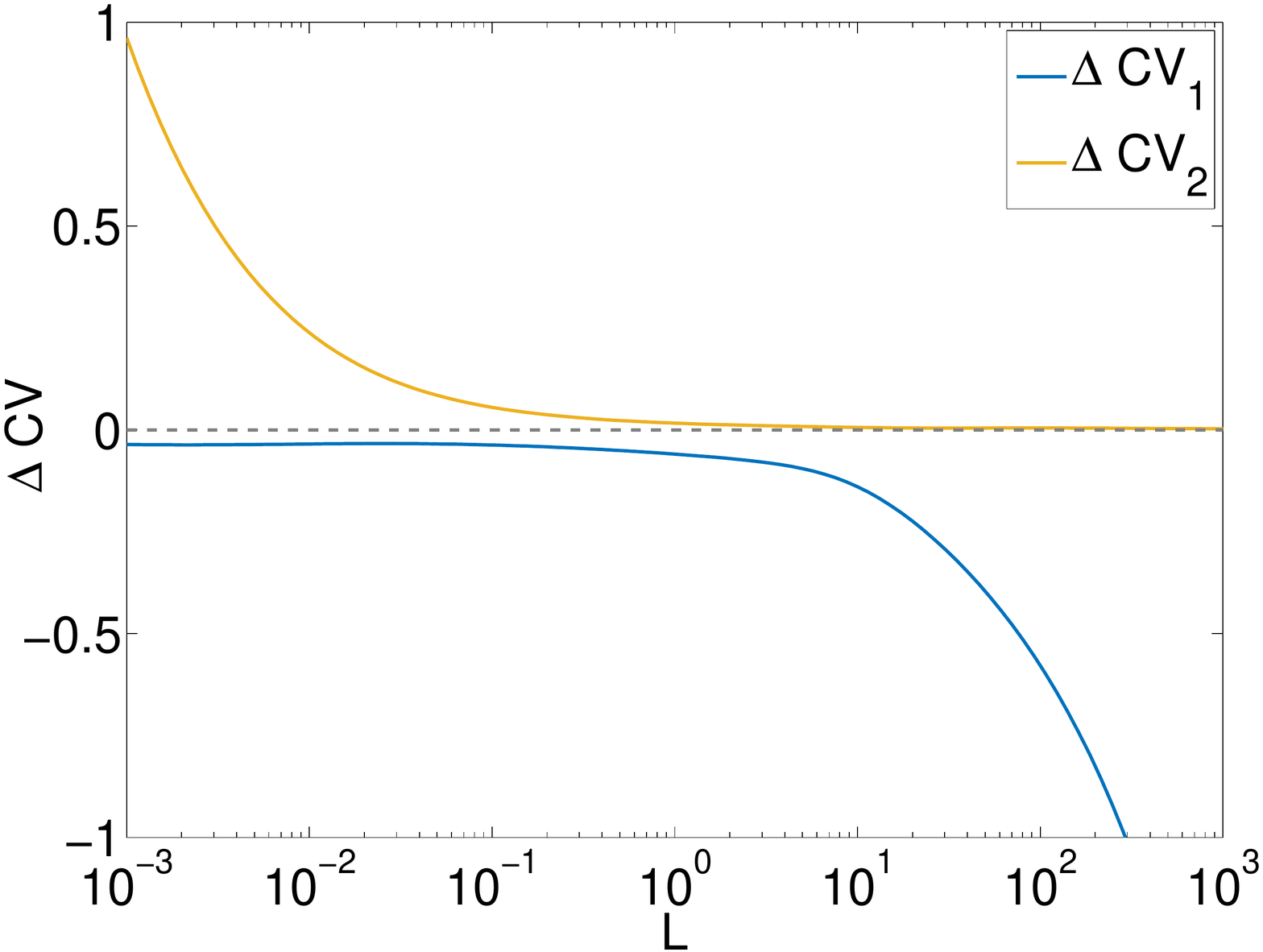}
}
\subfigure[\ ]{
\includegraphics[width=65mm]{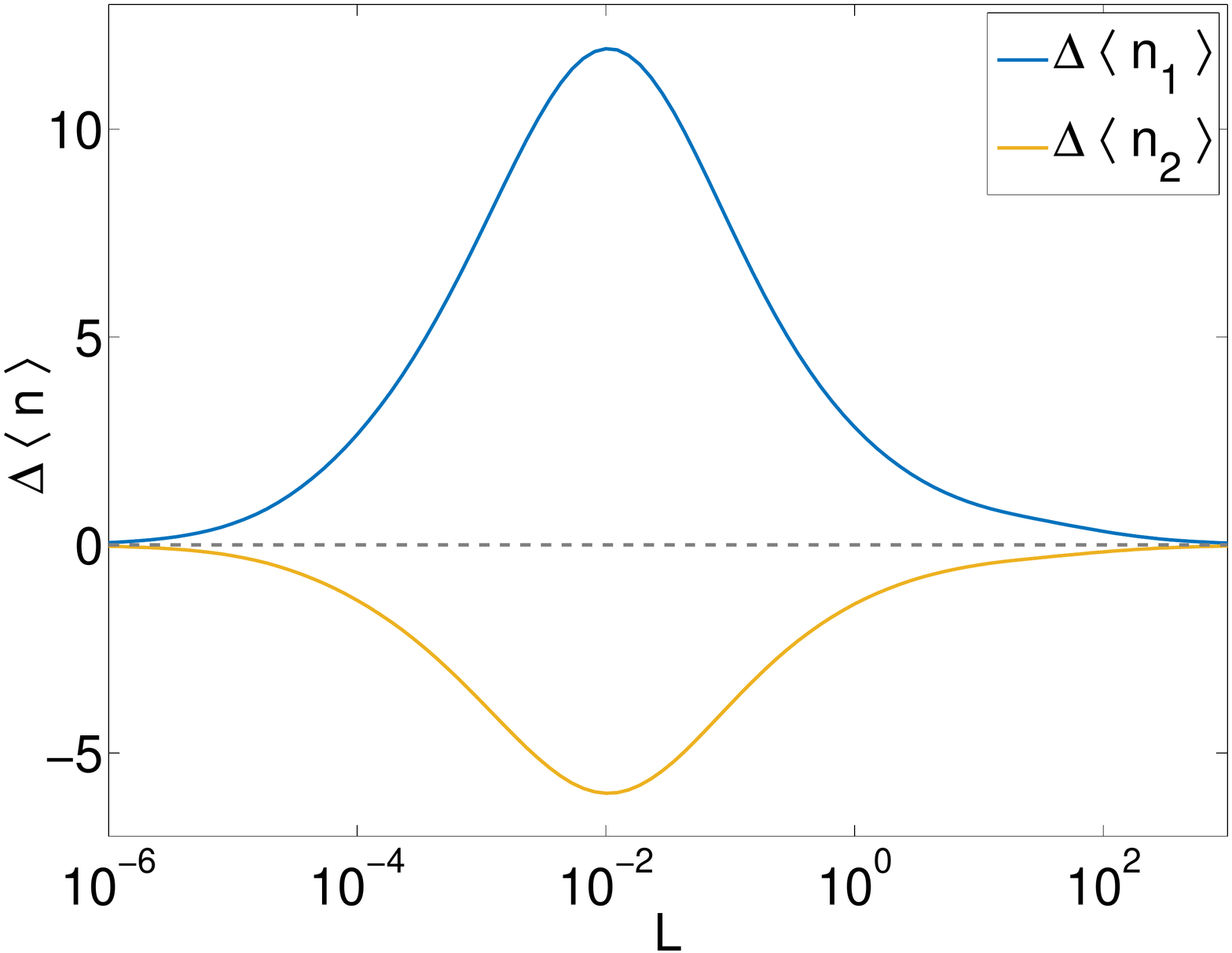}
}
\caption{Comparison of statistics of intrinsic noise in high volume exclusion and dilute conditions for the closed reversible dimerisation reaction. The statistics are all numerically calculated from Eq. (\ref{vCMEcd}); the dilute case is obtained by choosing $R = N/k = 1000$ and the high volume exclusion case by choosing $R = 1$. The constant $k$ is fixed to 50 in all cases. The non-dimensional parameter $L$ which is an aggregate of all rate constants and the volume is varied and the statistics plotted as a function of $L$. In (a) and (b) we show the variation of the Fano factors of the two species. In (c) we compute the difference between the CV in dilute and high volume exclusion conditions, $\Delta CV_i$, for both species. In (d) we compute the difference between the mean number of molecules in dilute and high volume exclusion conditions, $\Delta \langle n_i \rangle$. See text for discussion.
}
\end{figure}

\section{Comparison of the \lowercase{v}RDME with Brownian dynamics}
The vRDME has at least one major disadvantage -- it is based on an artificial spatial lattice. Ideally one would like to derive the vRDME rigorously from a lattice-free approach or at least to show that it is a reliable approximation of a lattice-free description under some conditions. 
\\\\
A derivation of this type has been previously attempted for the dilute case. In particular it has been shown that for systems with bimolecular reactions, the RDME provides a good approximation to the lattice-free descriptions offered by the Doi \cite{Doi1976a,Doi1976b} and Smoluchowski models \cite{Smoluchowski1917,Keizer1982} for lattice spacings that are neither too small nor too large \cite{Isaacson2009a}. In the limit of small lattice spacing, the statistics of the RDME do not converge to those of the lattice-free approach \cite{Isaacson2009,Hellander2012} but it is possible in this case to derive a new convergent RDME called the CRDME which does not suffer from this issue \cite{Isaacson2013}. 
\\\\
The question of agreement between a lattice and lattice-free approach in the case of volume excluded interactions is relatively simpler than for the dilute case because there is one less parameter: unlike the RDME, the lattice spacing of the vRDME is fixed to equal the molecule diameter. A formal derivation of the vRDME from the volume excluded versions of the spatially continuous Doi or Smoluchowski models is beyond the scope of this paper; here we shall be content with comparing the statistics of the vRDME with those obtained from microscopic Brownian dynamics (BD) simulations for a simple example. 
\\\\
In particular we test the validity of the RDME and vRDME by comparing their global distribution solutions for the closed dimerisation system \eqref{closeddimr} given by Eqs. \eqref{CMEcd} and \eqref{vCMEcd} respectively, with the distributions calculated from ensemble averaging BD simulations of the same chemical system. The BD simulations consider particles to be two-dimensional hard disks which move randomly in space according to standard Brownian motion. With parameters defined as in Eqs. \eqref{CMEcd} and \eqref{vCMEcd}, diffusion coefficient $D$ and time-step $\Delta t$, the BD algorithm we use is as follows:
\\\\
(1) Place $k$ particles of type $X_1$ with radius $r$ at uniformly distributed points in $[0,1] \times [0,1]$ such that they do not overlap. Set time $t=0$. Proceed to (2).
\\\\
(2) Propose a Normal random number with mean $0$ and standard deviation $\sqrt{2D\Delta t}$ to add to each particle coordinate. If no pairs of particles will overlap, accept the new coordinates and proceed to (4). If exactly one pair of particles will overlap and they are both type $X_1$, proceed to (3). Else reject the new coordinates and attempt (2) again. 
\\\\
(3) Choose a uniform random number \emph{rand} between 0 and 1. If \emph{rand} $\geq p\Delta t$ reject the new coordinates from (2) and attempt (2) again. Else if \emph{rand} $<p\Delta t$, remove the overlapping $X_1$ particles. Place a $X_2$ particle with radius $r$ midway between the centres of the removed particles. Choose an Exponential random number \emph{exprand} with mean $1/k_1$. Assign a number $\tau=t+$\emph{exprand} to the new $X_2$ particle. Proceed to (4).
\\\\
(4) For each $X_2$ particle, check if $t>\tau$. If not, proceed to (5). Else for each $X_2$ particle with $t>\tau$, remove it and place two $X_1$ particles just touching at a random angle such that their midpoint is the former centre of the $X_2$ particle. If any of the new $X_1$ particles overlap other particles, remove them, replace the $X_2$ particle, and set $\tau=t+$\emph{exprand}. Proceed to (5).
\\\\
(5) Advance time by setting $t=t+\Delta t$. Store the total number of $X_1$ and $X_2$ particles in memory. Return to (2) and repeat until a given time has elapsed. 
\\\\
Note that, in the above algorithm $p=\frac{k_0}{2\pi r^2}$ which is the probability per unit time that a given pair of $X_1$ particles reacts. This choice guarantees that in the limit of well-mixed and dilute conditions, the rate at which dimerisation occurs in the Brownian dynamics agrees with that given by the bimolecular propensity in the CME (for a derivation see Appendix D of \cite{Smith2015}). Note also that the precise choice of the distance at which one places the two particles of type $X_1$ when a dimer $X_2$ dissociates has little effect on the statistics collected, as long as we have reaction-limited kinetics (probability of the association of two particles of type $X_1$ is very small). The above algorithm can be considered a volume-excluded version of standard BD algorithms used for dilute reversible systems \cite{Lipkova2011}.
\\\\
For accuracy $\Delta t$ should be chosen small enough such that at most one reaction normally happens in each time step. To compare BD and vRDME, we choose the particles to have a diameter equal to the width of a vRDME voxel. This implies that the proportion of volume occupied by a BD particle is slightly less than the proportion of volume occupied by a vRDME voxel, however it is the most natural way of assigning a diameter, and it ensures that BD can feasibly reach the levels of volume exclusion that we want to model with the vRDME.

\begin{figure} [h]
\centering
\subfigure[\ Global distribution (low volume exclusion)]{
\includegraphics[width=65mm]{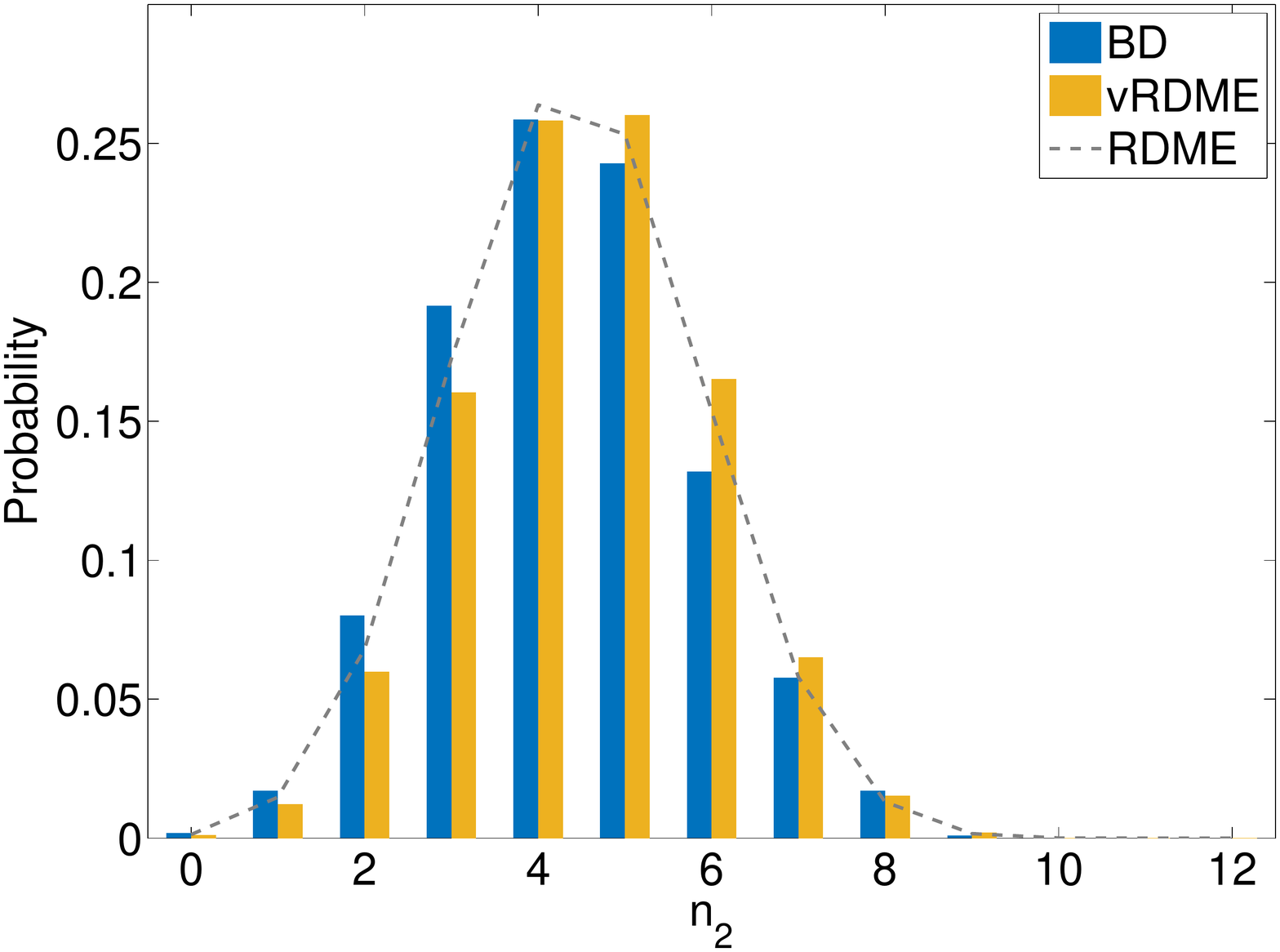}
\label{fig8subfig1}
}
\subfigure[\ Global distribution (high volume exclusion)]{
\includegraphics[width=65mm]{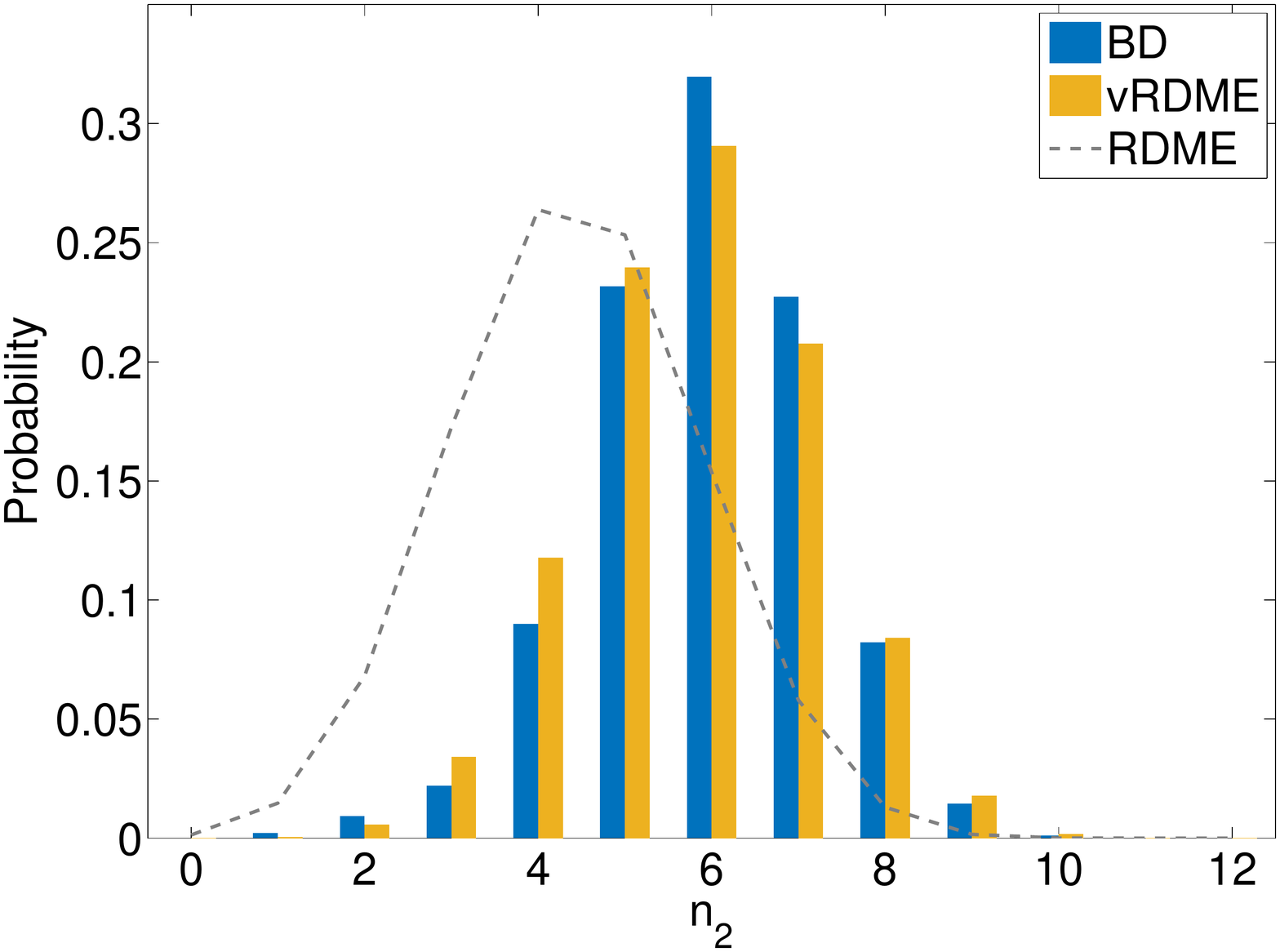}
\label{fig8subfig2}
}
\caption{(a,b) Comparison of exact vRDME and RDME distributions with BD distributions of molecule numbers of species $X_2$ in the closed dimerisation system. (a) The system with $3-6\%$ occupied volume, (b) the system with $33-67\%$ occupied volume. Parameter values are $k_0=0.01$, $k_1=0.5$, $\Omega=1$, $D=10^{-4}$, $\Delta t=10^{-2}$, (a) diameter=$\frac{1}{20}$, $N=400$, (b) diameter=$\frac{1}{6}$, $N=36$.}\label{fig8}
 \end{figure}

In Fig. \ref{fig8} we compare BD simulations with the exact global distributions of the RDME and the vRDME as given by Eqs. \eqref{CMEcd} and \eqref{vCMEcd} respectively. In panel \ref{fig8subfig1}, we show the equilibrium global probability distribution of $X_2$ computed with BD (blue histogram), vRDME (yellow histogram) and RDME (grey dashed line), in dilute conditions. In this case, in BD, the particle diameters were $\frac{1}{20}$ and there were 24 $X_1$ particles initially; equivalently, in the vRDME, the number of voxels is $N=400$. It follows that the percentage of occupied volume in this case varies from $3-6\%$, where $3\%$ is reached when all $X_1$ particles are bound in dimers $X_2$. Since this corresponds to fairly dilute conditions, it is unsurprising that BD, the vRDME and the RDME essentially agree. In panel \ref{fig8subfig2}, we show the same plot in high volume exclusion conditions. In this case, in BD, the particle diameters were $\frac{1}{6}$ and there were 24 $X_1$ particles initially; equivalently, in the vRDME, the number of voxels is $N=36$. Therefore the percentage of occupied volume in this case varies from $33-67\%$. Thus this corresponds to considerably high volume exclusion; the vRDME here agrees with BD but the RDME strongly disagrees with both. 
\\\\
\\\\
Hence our analysis confirms that for the dimerisation reaction, the vRDME gives global statistics that are in very good agreement with those obtained from a microscopic lattice-free approach, for a parameter set in both low and high volume exclusion scenarios. This is likely mainly due to the fact that the vRDME is a description at the natural length scale of the system (the molecular diameter). Further research is however necessary to clarify whether the agreement between the vRDME and BD holds for a broad range of parameter values and for general chemical systems.

\section{Summary and Conclusion}

In this paper, we have elucidated some of the effects which volume exclusion can have on intrinsic noise in chemical systems which are in equilibrium. In particular, the novelty of our study is that we can make precise statements on the relationship between the probability distribution solution of the master equation and the extent of volume exclusion. This was possible because we obtained an exact solution of the local and global probability distribution of the RDME and of its excluded volume version, the vRDME, in equilibrium (detailed balance) conditions. 

A summary of our findings is as follows. We found that the type of the global marginal distributions of the RDME and vRDME varies according to the type of chemical conservation law: (i) for those systems with no chemical conservation law, the global marginal distribution of the RDME and vRDME for all species is Poisson and Binomial respectively; (ii) for those systems with a chemical conservation law of the sum type, Eq. (\ref{conslaw1}), the global marginal distribution of the RDME and vRDME for a species not involved in the chemical conservation law is also Poisson and Binomial respectively; (iii) for those systems with a chemical conservation law of the sum type, Eq. (\ref{conslaw1}), the global marginal distribution of the RDME and vRDME for a species involved in the chemical conservation law is Binomial. Taking into account volume exclusion has very little or no impact on the fluctuations in this case; (iv) for those systems with a chemical conservation law of a more general type, nothing can be directly deduced about the type of marginal distributions because of the complexity of the exact normalised probability distributions. However for a specific system of this type we found that the global fluctuations were neither Poisson nor Binomial for species involved in the chemical conservation law and that volume exclusion did have a strong impact on the fluctuations, in contrast to systems with a chemical conservation law of the sum type. 

Given points (i)-(iii) above, we can clearly state that the largest impact of volume exclusion is likely to be on the intrinsic noise statistics of those species not involved in chemical conservation laws; the fact that the RDME solution is Poisson while the vRDME solution is Binomial implies that as the extent of molecular crowding increases, the fluctuations become increasingly sub-Poissonian, deviations from the classical inverse square root law for the noise-strength become more apparent and the marginal distribution of molecule number fluctuations changes from being skewed to the right (positive skewness) to being skewed to the left (negative skewness).

We note that the vRDME used in our study is based on an inherent assumption that the size of all molecules, reactant and inert, are roughly the same and equal to the size of a voxel. This is, of course, a gross simplification of reality, nevertheless the major benefits of this formulation is that (i) the vRDME is exactly solvable in equilibrium conditions and (ii) it appears to be an accurate approximation of microscopic spatially continuous stochastic simulations. Hence a comparison of the exact solution of the vRDME with the exact solution of the RDME (which assumes point particles) allows us to obtain a rough picture of the effects of volume exclusion on intrinsic noise, results which are difficult to obtain if we had to resort to computationally expensive stochastic simulations.  

Open questions which remain to be addressed involve understanding the impact of volume exclusion on non-equilibrium steady-states and on the time evolution of moments; these are challenging questions given that exact solutions of master equations are highly unlikely to be found in such conditions. Finally we expect the extension of the vRDME framework to allow the modelling of chemical reactions involving hard molecules of various sizes to be of paramount importance for the accurate prediction of the effect of volume exclusion on real chemical systems. 

\section*{Acknowledgments}

This work was supported by the BBSRC EASTBIO PhD studentship to S. S. and by a Leverhulme grant award to R. G. (RPG-2013-171). C. C. thanks Philipp Thomas for useful discussions.

\appendix

\section{Derivation of the global distribution of molecule numbers of the vRDME and RDME}

As discussed in Section III in the main text, it is straightforward to show that the solution of the vRDME in equilibrium conditions is:
\begin{align}
\label{AvRDMElocal}
P(n_1^1,...,n_{M+1}^1,...,n_1^N,...,&n_{M+1}^N) = \nonumber \\ &C  \prod_{k=1}^{N} \prod_{i=1}^{M+1}  \frac{((\Omega / N) \tilde{\phi}_i)^{n_i^k}}{n_i^k!} \delta(\sum_{i=1}^{M+1} n_i^k, 1) \prod_{m=1}^S \delta(f_m(n_1,n_2,...,n_M), K_m),
\end{align}
where $n_i$ is the global concentration of species $X_i$, i.e., $n_i = \sum_{j=1}^N n_i^j$. 

We now use Eq. (\ref{AvRDMElocal}) to calculate the probability over the vector of the global number of molecules $\vec{n}=\{n_1,...,n_M\}$. We start by noting that the definition of the global concentration of species $X_i$, i.e., $n_i = \sum_{j=1}^N n_i^j$ together with the conservation law factor $\prod_{k=1}^{N} \delta(\sum_{i=1}^{M+1} n_i^k, 1)$ is equivalent to the factor $\delta(\sum_{i=1}^{M+1} n_i, N)$. Thus we have:
\begin{align}
\label{AvRDMEglobal1}
P&(n_1,...,n_{M+1}) = C \sum_{n_i^k} P(n_1^1,...,n_{M+1}^1,n_1^2,...,n_{M+1}^2,....,n_1^N,...,n_{M+1}^N) \delta(\sum_{r=1}^N n_i^r,n_i), \\ \
\label{AvRDMEglobal2} &= C \biggl[ \sum_{n_i^k} \prod_{i=1}^{M+1}  \prod_{k=1}^{N} \frac{((\Omega / N) \tilde{\phi}_i)^{n_i^k}}{n_i^k!}  \delta(\sum_{r=1}^N n_i^r,n_i) \biggr] \delta(\sum_{i=1}^{M+1} n_i, N) \prod_{m=1}^S \delta(f_m(n_1,n_2,...,n_M), K_m), \\
\label{AvRDMEglobal3} &= \prod_{i=1}^{M+1}  C \frac{(\Omega \tilde{\phi}_i)^{n_i}}{n_i!} \delta(\sum_{i=1}^{M+1} n_i, N) \prod_{m=1}^S \delta(f_m(n_1,n_2,...,n_M), K_m).
\end{align}
The passage from Eq. (\ref{AvRDMEglobal1}) to Eq. (\ref{AvRDMEglobal3}) can be explained as follows. The sum in Eq. (\ref{AvRDMEglobal1}) is over the local molecule numbers only and hence the delta function over the global molecule numbers $ \delta(\sum_{i=1}^{M+1} n_i, N) \delta(f_m(n_1,n_2,...,n_M), K_m)$ are unaffected by this sum and can be left outside, which leads to Eq. (\ref{AvRDMEglobal2}). Now the term in square brackets in the latter equation is a product of independent Poissonians (the correlation between Poissonians is induced by the delta functions outside of the square brackets). Due to the delta function $\delta(\sum_{r=1}^N n_i^r,n_i)$, the sum in the square brackets amounts to calculating the probability distribution of a sum of independent Poisson random variables, which leads to the final result Eq. (\ref{AvRDMEglobal3}). Note that Eq. (\ref{AvRDMEglobal3}) is the same as the equilibrium solution of the vCME, Eq. (\ref{vCMEsol}), which establishes the equivalence of the vRDME and vCME at the global level in equilibrium conditions. By an analogous approach, one can also show the equivalence of the RDME and CME at the global level in equilibrium conditions.

 \end{document}